\documentclass[preprint2]{aastex631}

\shorttitle{Kepler's Greatest Hits}
\shortauthors{Dattilo et al.}
\graphicspath{{./}{figures/}}
\usepackage{xcolor}
\usepackage{amsmath}

\newcommand{\Kepler}{\textit{Kepler}}
\renewcommand{\Re}{R$_\oplus$}
\defcitealias{kunimoto2020b}{KM20}
\defcitealias{petigura2018}{P18}
\defcitealias{hsu2019}{H19}

\usepackage{etoolbox}
\makeatletter
\patchcmd{\ltx@foottext}{%
  .5\textwidth\advance\hsize-18pt}{%
  \linewidth\advance\hsize-1.8em%
}{}{}
\makeatother

\begin{document}

\title{A Unified Treatment of Kepler Occurrence to Trace Planet Evolution I: Methodology}

\author[0000-0002-1092-2995]{Anne Dattilo}
\affiliation{Department of Astronomy and Astrophysics, University of California Santa Cruz, Santa Cruz, CA, 95064, USA}
\author[0000-0002-7030-9519]{Natalie M. Batalha}
\affiliation{Department of Astronomy and Astrophysics, University of California Santa Cruz, Santa Cruz, CA, 95064, USA}
\author[0000-0003-0081-1797]{Steve Bryson}
\affiliation{NASA Ames Research Center, Moffett Field, CA 94035, USA}

\correspondingauthor{Anne Dattilo}
\email{adattilo@ucsc.edu}

\begin{abstract}
We present \Kepler{} exoplanet occurrence rates for planets between $0.5-16$ \Re{} and between $1-400$ days. To measure occurrence, we use a non-parametric method via a kernel density estimator and use bootstrap random sampling for uncertainty estimation. We use a full characterization of completeness and reliability measurements from the \Kepler{} DR25 catalog, including detection efficiency, vetting completeness, astrophysical- and false alarm reliability. We also include more accurate and homogeneous stellar radii from \textit{Gaia} DR2. 
In order to see the impact of these final \Kepler{} properties, we revisit benchmark exoplanet occurrence rate measurements from the literature.
We compare our measurements with previous studies to both validate our method and observe the dependence of these benchmarks on updated stellar and planet properties.
For FGK stars, between $0.5-16$ \Re{} and between $1-400$ days, we find an occurrence of $1.52\pm0.08$ planets per star. We investigate the dependence of occurrence as a function of radius, orbital period, and stellar type and compare with previous studies with excellent agreement. We measure the minimum of the radius valley to be $1.78^{+0.14}_{-0.16}$ \Re{} for FGK stars and find it to move to smaller radii for cooler stars. We also present new measurements of the slope of the occurrence cliff at $3-4$ \Re{}, and find that the cliff becomes less steep at long orbital period. Our methodology will enable us to constrain theoretical models of planet formation and evolution in the future.
\end{abstract}

\section{Introduction} \label{sec:intro}
\Kepler{}'s foremost legacy has been enabling detailed exoplanet demographic studies. Launched in 2009, \Kepler{} was designed as a demographics mission to study the population of planets in our galaxy orbiting within 1 au of their host stars \citep{borucki2010}. It also aimed to measure the frequency of Earth-like planets within their star's habitable zone. These small planets are difficult to detect around Sun-like stars due to both their small size and long orbital period. To detect these planets, \Kepler{} observed a single field continuously for almost four years.

In order to enable studies of planetary demographics, the \Kepler{} Mission built a homogeneous planet catalog. A considerable amount of resources went into characterizing this catalog: both in the completeness (the fraction of planets that were correctly identified and vetted as planet candidates) and in the reliability (the fraction of planet candidates that are truly planets). The final uniform planet catalog, Data Release 25 (hereafter DR25), was released in 2018 and contained 4,043 planet candidates \citep{thompson2018}.

Over the past decade, a large amount of work has gone into measuring \Kepler{} occurrence rates. 
Many early works \citep[e.g.][]{borucki2011, catanzarite2011, youdin2011, howard2012, traub2012} aimed to measure the frequency of small planets with the first two \Kepler{} planet catalog releases, which contained a cumulative 1,235 planets found in the first 13 months of data \citep{borucki2010, borucki2011}. In the following years, multiple intermediate planet catalogs were created, and many occurrence rate studies were performed on them \citep[e.g.][amongst others]{mann2012, wright2012, swift2013, dressing2013, fressin2013, beauge2013, petigura2013, kopparapu2013, dong2013, kane2014, dfm2014, morton2014, silburt2015, mulders2015, burke2015, dressing2015, bryan2016, mulders2016, gaidos2016, narang2018, zhu2018, khu2019}.

These studies were improved with an accurate model of the \Kepler{} completeness function. Understanding the \Kepler{} completeness function via injection and recovery tests \citep{christiansen2013, christiansen2015, christiansen2016, christiansen2017, burke2017} proved successful in improving our understanding of the intrinsic planet population. \citet{burke2015} first used direct measurements of the \Kepler{} detection efficiency to present occurrence rate calculations on the Q1-16 \Kepler{} planet candidate sample from \citet{mullally2016}.
These improvements revealed new features in the period-radius plane: \citet{fulton2017} uncovered the radius valley, long theorized but not previously observationally seen. This feature is robust against various stellar samples. It can be seen in \textit{K2} data \citep{khu2019}, implying its formation cause is not reliant on galactic position. Further, \citet{vaneylan2018} showed that with asteroseismology, the valley emptied, indicating that the more precise our stellar parameters, the more features we can pull out of the true planet population.

However, few occurrence studies used the DR25 catalog with DR25 completeness measurements, and with few exceptions, these measurements also did not incorporate any reliability measurements for individual planets. \citet[][hereafter H19]{hsu2019} calculated \Kepler{} occurrence rates with DR25 and updated, \textit{Gaia}-derived stellar properties,  however without any reliability measurements. \citet[][hereafter KM20]{kunimoto2020b} used catalog reliability in their measurements but used their own pipeline to detect planets from the Q1-Q17 data \citep{kunimoto2020a}. \citet{bryson2020a} calculated \Kepler{} occurrence rates using a joint DR25-\textit{Gaia} DR2 stellar catalog as well as DR25 reliability measurements, albeit for a focused domain in the period-radius domain.

Precise and uniform planet catalogs allowed us to measure new structures in the period-radius diagram, just as they did to reveal the radius valley in \citet{fulton2017}. The period radius diagram is rife with features, such as the Neptune desert at low orbital period, the aforementioned radius valley bifurcating the small planet population, and the radius cliff that marks the steep drop in occurrence from the sub-Neptunes to the Neptunes. Many of these structures in the period-radius diagram have been tied to planet formation and evolution theory \citep[e.g.][]{mazeh2016, ginzburg2016, owen2017, owen2018, kite2019}. Billions of years of planet formation have driven our planets to their current places on the period-radius diagram. We can trace the history of planet formation through these end locations. By measuring planet occurrence, we can map the structure of the period-radius diagram to the physics that put each planet there.
Therefore, the better we can describe the underlying structure of occurrence in the period-radius diagram, the better we can explain how planets form and evolve.

The purpose of this paper is to revisit the planet occurrence rate measurements from the past decade, now with a uniform treatment of all DR25 inputs, in order to well-measure the structure in occurrence space. We develop a methodology to calculate planet occurrence using a kernel density estimator. Our methodology incorporates DR25 completeness and reliability measurements in a non-parametric form in order to measure planet occurrence rates. 
Because it is of particular interest to measure structure in the period-radius diagram, the future goal of using this KDE methodology is to glean complex structure in multiple dimensions from the data, instead of fitting parametric forms to it. To validate the method, we have identified several benchmark demographic measurements from previous \Kepler{} studies: the hot Jupiter occurrence, the radius distribution for small planets, and a parameterized period distribution.

Due to evolving planet catalogs and stellar samples, there has been a spread in the measurements for each of these benchmarks over the years. We measure these with our methodology and compare to previous studies, but we are unable to make truly direct comparisons between our measurements and those of previous studies because of different input assumptions.
We combine the final DR25 parameters with \citet{berger2020a} stellar parameters so we can see how adding in the updated stellar values and reliability measurements impact the final results.
 
The paper is structured as follows. In Section~\ref{sec:stars} we discuss the stellar and planet samples we use from the \Kepler{} survey. We then discuss how we characterize the survey bias with completeness and reliability measurements, as well as our methodology for measuring planet occurrence in Section~\ref{sec:methods}. In Section~\ref{sec:occ} we compare to other works and measure features on the period-radius diagram. In Section~\ref{sec:discussion} we tie these rates to physical theory. We conclude and lay out possible future work in Section \ref{sec:conclusions}.

\section{Stellar Sample} \label{sec:stars}
We base our parent stellar sample on the Q1-17 DR25 stellar catalog, containing 198,709 stars \citep{mathur2017}. In order to have the most precise and uniform catalog of stellar properties, we use \citet{berger2020a} stellar properties. \citet{berger2020a} derives a homogeneous catalog of stellar properties from isochrone fitting using \textit{Gaia} parallaxes, broadband photometry, and spectroscopic metallicities where available. The \citet{berger2020a} stellar properties cut down on the dominant sources of systematic errors in stellar radii by deriving them from isochrone-fit temperatures and luminosities, resulting in the average stellar radii error of less than 10\%. Because all stellar parameters used in our model are from this catalog, we restrict the stellar sample in the DR25 catalog to those included in \citet{berger2020a}, cutting the stellar sample down to 177,661 stars.

We then make further cuts to our stellar catalog following \citet{bryson2020a}. First, we make cuts for binarity. Stellar multiplicity can bias measured planet properties and suppress planet occurrence for close-in binaries \citep{sullivan2022}. The Gaia Renormalized Unit Weight Error (RUWE) is often used as a stellar multiplicity indicator. It is a goodness-of-fit metric to the astrometric fits based on \citet{lindegren2018}, who suggest a single-star cut-off at RUWE $>$ 1.4.
We apply a slightly harsher cut of 1.2 because there are few single stars below this limit \citep[see e.g.][for more discussion]{bryson2020a, berger2020a}. After removing targets with RUWE $>$ 1.2, there are 163,254 stars remaining.
\textit{Gaia} DR3 contains additional flags for binarity, specifically the \texttt{non\_single\_star} flag. We did not consider this flag prior to our analysis. The RUWE cutoff, however, is very effective, removing all but one of the 2319 binaries identified by the \texttt{non\_single\_star} multiplicity flag. The remaining flagged star has a RUWE value below the ``typical'' cutoff of 1.2, and therefore would not have been cut with a harsher RUWE metric. This single exception does not justify the cost of re-analysis and it remains in our stellar sample.
We also remove stars that are flagged in \citet{berger2018} as likely binary (BIN flag = 1 or 3). Stars were flagged based on their inflated radii in the H-R diagram (161,075 remain). We leave stars flagged with BIN = 2 because the flag comes from high-resolution imaging, which was only performed for a subset of stars.
We remove stars with \citet{berger2020a} goodness-of-fit parameter $<$ 0.99, which measures the quality of isochrone fitting (160,638 remain).
Our final stellar classification cut is to remove stars flagged as evolved stars according to \citet{berger2018}. This flag is determined using solar-metallicity \texttt{PARSEC} evolutionary tracks \citep{bressan2012} to define the terminal age main sequence stars in the temperature-radius plane; anything above this boundary is classified as a subgiant or red giant star (and cut from our sample; 107,472 remain).

We then cull our sample of poor targets. We remove noisy targets identified in the KeplerPORTs package\footnote{https://github.com/nasa/KeplerPORTs} (105,994 remain) and stars with NaN limb darkening coefficients (105,538 remain). We then cull based on duty cycle, the fraction of data cadences with valid data: we check for stars with duty cycle $=$ NaN (105,538 remain), targets with duty cycle drop $> 0.3$ (101,544 remain), and targets with duty cycles $< 0.6$ (98,150 remain). We remove stars with data span $<$ 1000 days, which ensures that each star in the sample has enough data to meet the 3-transit requirement for orbital periods of 500 days (89,868 remain). The cut removes some stellar targets with short period planets, but we require a uniform treatment for all stellar targets. Lastly, we remove stars with the \texttt{timeoutsumry} flag $!=$~1, which removes incompletely searched targets, resulting in 84,341 stars.

We split our sample by stellar type using the temperature boundaries from \citet{pecaut2013}. The remaining 84,341 stars are thus split into 21,999 F stars $(6000$ K $\leq T_{eff} < 7300$ K); 41,501 G stars $(5300$ K $\leq T_{eff} < 6000$K); 19,412 K stars $(3900$ K $\leq T_{eff} < 5300$ K). We use the sample of 82,912 FGK stars for our parent sample in analysis.

\subsection{Planet Sample}\label{sec:planets}
We derive our planet sample from the Q1-17 DR25 KOI table, restricted to KOIs with \texttt{koi\_disposition} = CANDIDATE on the stars in our parent stellar sample. We recompute the planet radius, $R_P$, based on the ratio of planet radius to stellar radius, $R_P / R_*$ (\texttt{koi\_ror}), using the updated stellar radii from \citet{berger2020a}. Because there is an excess false positive rate below 1 day in both the TCE population and the PC sample (see figure 1 and 7 of \citet{thompson2018}), we cut our planet sample at 1 day. 

\begin{figure}
    \centering
    \includegraphics[width=\columnwidth]{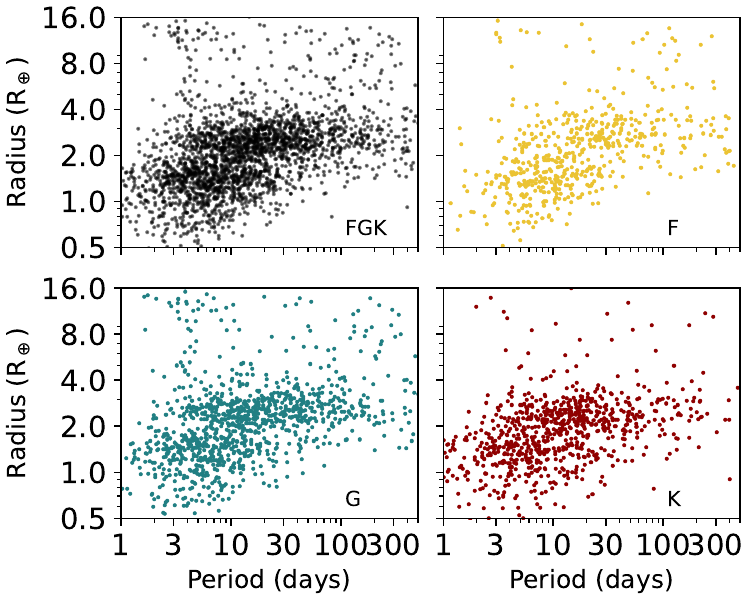}
    \caption{Period-radius diagram of the 2,575 planet candidates in our planet sample, divided by stellar type.}
    \label{fig:planets}
\end{figure}

After culling the stellar sample and cutting on planet orbital period and radius, we are left with 2,575 planet candidates in our total FGK sample. There are 610 planet candidates around F stars, 1275 planet candidates around G stars, and 690 planet candidates around K stars.
Figure~\ref{fig:planets} shows the period-radius distribution of the planet sample.

\section{Methodology} \label{sec:methods}
With an idealized (100\% complete and reliable) dataset, our definition of planet occurrence for a defined stellar sample would be

\begin{equation}
    \textrm{NPPS} = \frac{N_P}{N_*}
\end{equation}\\
where the number of planets per star (NPPS) is the number of detected planets, $N_P$, divided by the total number of stars $N_*$. In practice, however, not all planets are detected by our survey, nor are all the planets our pipeline detects truly planets. We can attempt to account for these biases by computing various performance metrics for the survey: we must account for both the \textit{completeness} of our survey and the \textit{reliability} of our sample.

The \Kepler{} pipeline identifies potential transit signals in lightcurves as threshold crossing events, or TCEs \citep{jenkins2002}. The Robovetter, an automated vetting algorithm that dispositions signals based on decision trees \citep{thompson2015}, classifies these events as planet candidates (PCs) or not. These signals are not always correctly categorized, however. Some planets are missed, and other dispositioned PCs are not actual planets.
Completeness (or sensitivity, or recall) asks: out of all real planets, how many do we correctly disposition as planet candidates? Reliability (or precision) asks, out of all TCEs dispositioned as planet candidates, how many are true positives? 

Using measured completeness and reliability, we update our occurrence equation:

\begin{equation}\label{eq:npps}
    \textrm{NPPS} = \frac{1}{N_*}\sum_{i}^{}{w_i}
\end{equation}\\
where we sum over all detections, using $w_i$ as our weighting factor to account for the bias in our survey. The weighting factor $w_i$ is defined as:
\begin{equation}\label{eq:idem}
    w_i = \frac{r_i}{\omega_i}
\end{equation}\\
$\omega$ is total summed completeness, which can be calculated at each ($p_i$, $r_i$) for each planet candidate as defined in Section~\ref{sec:completness}, and $r$ is reliability, calculated for each planet as defined in Section~\ref{sec:rel}. The method expressed in Eq.~\ref{eq:idem} to calculate occurrence is generally called the ``Inverse Detection Efficiency Method'' or IDEM \citep{howard2012, petigura2013, dfm2014}. We build upon this method using a kernel density estimator (Section~\ref{sec:kde}).

\subsection{Completeness} \label{sec:completness}
Our completeness model is based on the \citet{bryson2020a} model and is comprised of two parts: detection and vetting completeness. 
Detection completeness is the fraction of true transiting planets that the \Kepler{} pipeline could detect. Vetting completeness is the fraction of true transiting planets that are classified as PCs by the vetting pipeline. We expect both to be a function of orbital period and the SNR of the transit light curve, which the \Kepler{} pipeline measures as the multiple event statistic (MES) \citep{jenkins2002}.

First, we calculate detection efficiency via a modified version of the KeplerPORTs code base.  
\citet{burke2017} used flux-level injection tests to generate detection efficiency curves for individual target stars, which are a function of expected MES. KeplerPORTs calculates an expected MES value for a hypothetical planet at any given period and radius within the given parameter space, resulting in a per-star completeness contour $\omega_s (p, r)$ as a function of period and radius. Geometric transit probability, the probability the planet can be seen transiting its host star, is also computed at this step.

We then compute vetting completeness. Vetting completeness is the fraction of detected TCEs that are correctly dispositioned as planet candidates (PCs). We use the same process to calculate vetting completeness as \citet{bryson2020a}, where it is described in detail, and briefly summarize it here.

Vetting completeness is computed by analyzing the fraction of recovered PCs in the injected set of planets from \citet{christiansen2017} in period-expected MES space. 
We grid period-expected MES space into cells and treat the fraction of correctly vetted PCs in each cell as a binomial rate. We fit these rates to a surface in period-expected MES space. 
As in \citet{bryson2020a}, this rate surface is the product of a non-rotated simplified logistic function in period and a rotated logistic in period and expected MES, fit via MCMC inference using \textit{emcee} \citep{emcee}. Since we use a different stellar sample than \citet{bryson2020a} (FGK vs. GK), we re-fit the parameters. We do this for each stellar type (F, G, and K separately) but find no substantial difference between stellar types. We do find differences between our FGK fit and the GK fit from \citet{bryson2020a}, which is expected for the differences in stellar samples. See Appendix~\ref{ap:vet} for the specific coefficients.
 
Each component of the completeness function (detection efficiency, geometric transit probability, and vetting completeness) is treated as independent; for each star, they are multiplied together to calculate total completeness. We sum these functions over all stars to get an average completeness over the total stellar sample.

We do not account for how transit multiplicity impacts our detection efficiency. When searching for PCs, the highest S/N signal is removed from the lightcurve by the \Kepler{} pipeline, leaving less and less data for each subsequent search. The detection efficiency for higher-order planets decreases as a result and is a function of both planet radius and orbital period \citep{zink2019}.
Properly accounting for this effect is out of the scope of this paper. However, we would expect it to decrease our completeness contours (we are not accounting for missing planets), which would increase our occurrence rates. When comparing parametric fits to measured occurrence, the GK occurrence rates computed by \citet{zink2019} agree with those of \citet{burke2015}, so we would expect the resulting difference in our measurements to be small.

\begin{figure*}
    \centering
    \includegraphics[width=\textwidth]{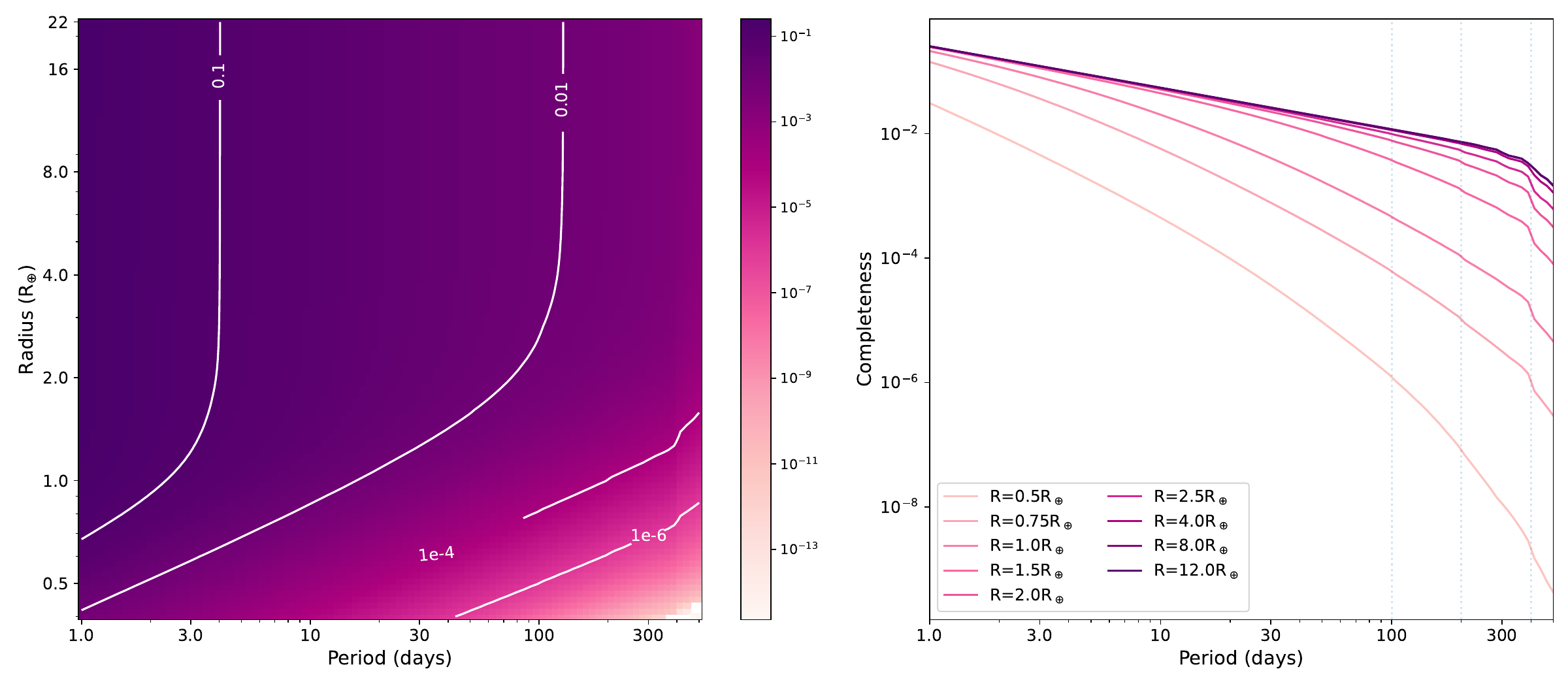}
    \caption{Left: FGK completeness function. Completeness degrades quickly along orbital period. White contours mark the 10-, 1-, 1e-2-, and 1e-4\% completeness levels. Right: the completeness contour for specific radii. There is a discontinuity in occurrence at 400 days caused by discontinuities in the detection efficiency model.}
    \label{fig:comp_contour}
\end{figure*}

The FGK completeness can be seen in Figure~\ref{fig:comp_contour}. It degrades as both a function of orbital period and planet radii and does not degrade smoothly. We would not expect it to be entirely smooth, though. 
For example, the window function is a key component to computing occurrence which quantifies the probability that a target star meets the minimum of three transits in observational data. It is not smooth due to gaps in observations. 
There is also noticeable structure that appears as jumps at exactly 100, 200, and 400 days. This is a result of a correction factor for orbital period in the detection efficiency model (see Section 3.4 of \citet{burke2017}) computed in discrete bins. The discontinuities due to the window function average out over our entire stellar sample, but orbital period discontinuities do not. Those remaining discontinuities can be seen in the right panel of Figure~\ref{fig:comp_contour}. Though small, they can add additional structure to our occurrence maps. For this study, we limit our occurrence measurements to 400 days.

\subsection{Reliability} \label{sec:rel}
In addition to completeness, we must also characterize the potential for false positives contaminating our planet sample via reliability measurements. Using the methodology of \citet{bryson2020a}, we use a probabilistic approach to characterize whether a planet candidate is actually a false positive. We break these into two categories: 1) false positives, which are due to astrophysical signals that imitate transits, and 2) false alarms, which are due to instrumental systematics or stellar activity. The process to calculate the different parts of reliability is described in detail in \citet{bryson2020a}, and its effects are characterized in-depth in \citet{bryson2020b}.

PCs can be false positives due to astrophysical causes, such as grazing or eclipsing binaries. We use the Q1-Q17 DR25 False Positive Probabilities Table for our false positive probabilities. These probabilities were created using the technique developed in \citet{morton2014} and were computed for all KOIs based largely on photometric data in \citet{morton2016}. We define the astrophysical reliability as $(1 \, -\, \textrm{the false positive probability})$. 

We characterize the false alarm reliability with an algorithm very similar to the vetting completeness. The inputs used to characterize are very different: instead of injected data, used to find how incomplete the pipeline is, we use inverted and scrambled data \citep{coughlin2017}. Inverted data inverts transit signals in lightcurves, while scrambled data mixes the lightcurve data order. Both of these create data that cannot contain true transits or astrophysical false positives, so any transit-like TCE is due to instrumental false alarms. As explained in \citet{thompson2018}, these inverted and scrambled false alarms have the same statistics as the dominant source of false alarms in the observed data.

The fraction of recovered signals in inverted and scrambled data is treated as a binomial rate problem in period-expected MES space, and MCMC inference is used to fit parameters to that surface. We fit the false alarm reliability with a rotated logistic function. As with vetting completeness, we fit each stellar type separately and find no substantial difference between the resulting posteriors. However, we do find a difference between the fit of \citet{bryson2020a} and our FGK fit. See Appendix~\ref{ap:rel} for the specific coefficients.

The total reliability for each planet candidate is the product of the false positive probability and the false alarm probability. 

\subsection{Computing Occurrence} \label{sec:kde}
The simplest way to measure occurrence is to count planets within some defined bin in period and radius (Eq.~\ref{eq:npps}). However, this binning can leave ``binning artifacts'' in the measured occurrence \citep{dfm2014}. One can also fit power law functions to the data to measure occurrence rates, but these fits limit the measured occurrence to that specific model, which may not be the correct choice. 

We use a kernel density estimator (KDE) to measure planet occurrence. KDEs allow us to measure closer to the underlying distribution by not assuming a model. We are not the first to employ KDEs in our occurrence framework. Other studies include \citet{morton2014, dressing2015, mulders2015c, jin2021, petigura2022}.

We define $f$ as the true planet occurrence and $\hat{f}$ as our measured occurrence. Traditionally, a KDE attempts to measure the underlying probability distribution of a sample using a function:
\begin{equation}
    \hat{f}(x) = \frac{1}{N} \sum_{i=1}^{N}{K(x-x_i; h)} 
\end{equation}

where $K$ is the kernel function and $h$ is the bandwidth. We use a 2D Gaussian kernel to represent the planet in log-uniform period-radius space, so $\hat{f}(x)$ becomes $\hat{f}(p, r)$. The bandwidth is the size of the kernel and acts as a smoothing function for the discrete dataset. The bandwidth needs to be optimized so that the final $\hat{f}$ is not over- or under-smoothed. For this paper, we use an empirically optimized bandwidth of $(\sigma_p, \sigma_r) = (0.05, 0.05)$ (see Section~\ref{sec:struc}).

Our resulting function is:
\begin{equation}
    \hat{f}(p, r) = \frac{1}{N} \sum_{i=1}^{N}{K(p-p_i; \sigma_p) K(r-r_i; \sigma_r)} 
\end{equation}

where $(p_i, r_i)$ represents each planet in our sample and $(\sigma_p, \sigma_r)$ denotes the bandwidth of (0.05, 0.05). 

In essence, we place each planet in our sample down at its period and radius as a bivariate Gaussian with width (0.05, 0.05) and amplitude equal to its reliability value, normalized so that the area under its curve is also equal to its reliability. We calculate the variance of $\hat{f}$ by performing a bootstrap of our sample 5000 times. 
For each bootstrap iteration, we calculate a new KDE, $\hat{f}^*$, and average the collection for our final KDE model. The collection of $\hat{f}^*$ measures the variance about $\hat{f}$, which mimics the variance of $\hat{f}$ about $f$. Our final occurrence rate KDE is computed by dividing $\hat{f}$ (which is equivalent to the averaged $\hat{f}^*$) by the completeness contour described in Section~\ref{sec:completness}. 

Because the KDE is a non-parametric method, we are not limited to a set grid for our measurements. We compute the KDE over a 1000x1000 grid and then integrate to smaller bin sizes such that there are detections in each bin. This gives us the flexibility to be able to change binning scheme to compare to other studies.

Throughout the paper, we refer to the average number of planets per star over a set domain as NPPS, which is computed as:
\begin{equation}
    \textrm{NPPS} = F = \int_P \int_R \frac{\hat{f}(p, r)}{\omega} \,dr\,dp
\end{equation}

\subsection{Method Validation}\label{sec:toy}
There are two questions we seek to address on the limitations of this methodology. Because the KDE kernel acts as a smoothing parameter, we need to understand what scale of structure we are able to measure. We also need to understand how we are limited by KDE edge effects and our survey completeness.

To address both of these, as well as to justify the choice of kernel size, we create a toy model. We generate a synthetic population with a known occurrence rate across our parameter space. Each (log-uniform) grid cell is sampled by a Poisson function, where the Poisson rate is the occurrence rate within that cell. This gives us the number of planets per cell. Each planet is then assigned a period and radius value from a uniform distribution bounded by the cell edges.

To create the \textit{detected} synthetic population, we must determine if each planet is ``observed.'' Whether a planet is observed or not can be thought of as a Bernoulli trial-- it is either observed (a success) or not (a failure). We set an observational completeness across the entire parameter space. The completeness value at each planet's period and radius acts as the probability of success for each ``trial.'' The planets successfully ``observed'' create the detected synthetic population, ready for our KDE pipeline. We use a reliability value of 1 for every planet. Once the KDE is created, it is divided by the completeness to get the measured occurrence. To validate the method, we compare input and measured occurrence. We input an occurrence of 1.0 planet per star within each grid cell. Between $1-400$ days and $0.5-16$ \Re{} we measure an average occurrence of $1.04\pm0.06$ . The occurrence outside this range, between $0.4-0.5$ \Re{} and $16-22$ \Re{}, is systematically lower due to KDE edge effects (discussed in Section\ref{sec:uniform}). We examine the level of structure we can measure in Section~\ref{sec:struc} and low completeness boundaries and KDE edge effects in Section~\ref{sec:uniform}.
 
\subsubsection{Testing Structure}\label{sec:struc}

\begin{figure*}
    \centering
    \includegraphics[width=\textwidth]{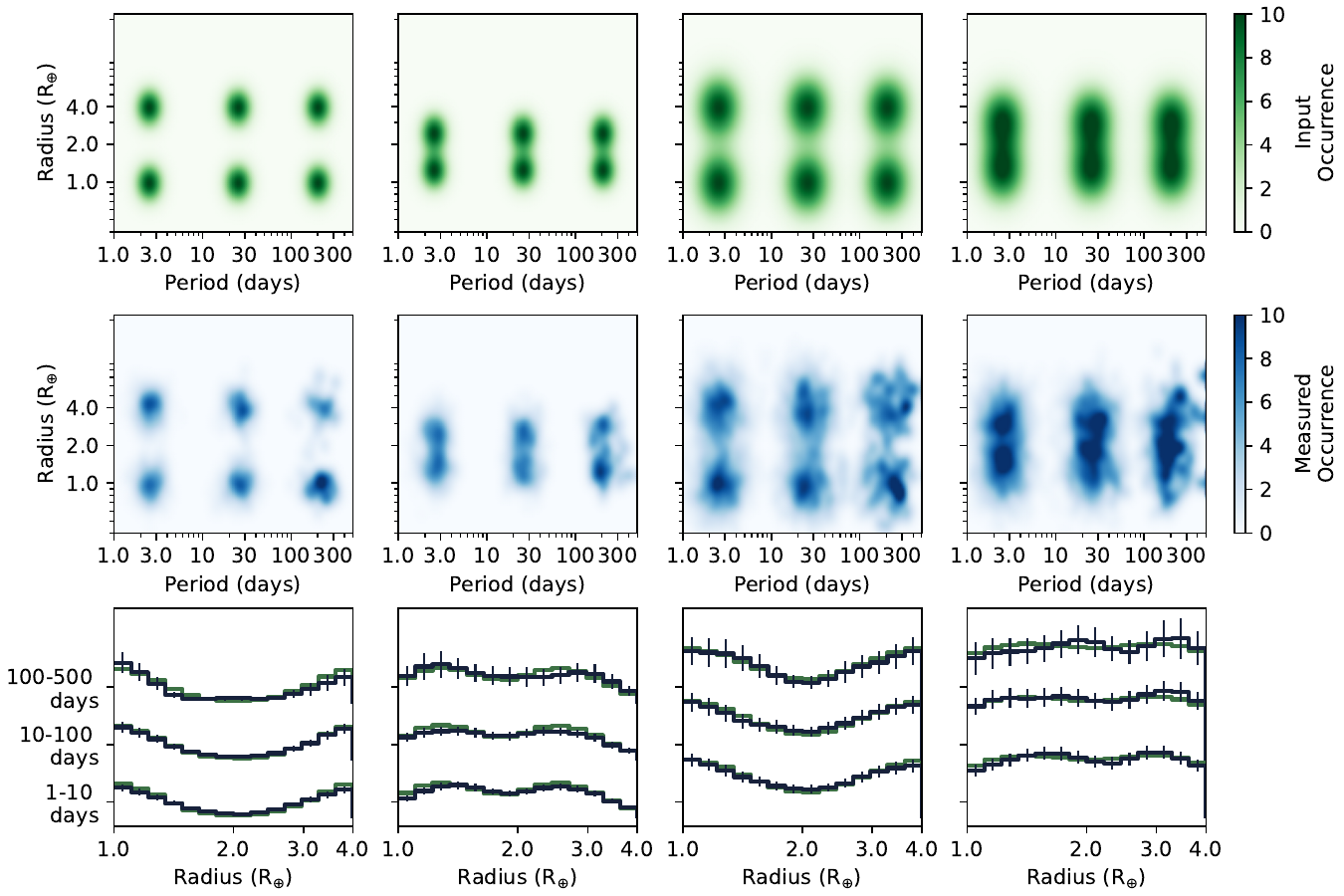}
    \caption{Testing what type of structure we can measure with a kernel size of $(\sigma_x, \sigma_y)=$(0.05, 0.05). From top to bottom, the three parts to test our ability to measure structure: the input occurrence (top, blue); the measured occurrence (middle, green); and the marginalized radius distributions for each pair of Gaussian occurrence, with the input occurrence in green and the measured occurrence in blue, with 1-$\sigma$ uncertainties. We test four different scenarios as described in the text. For each scenario, we use a linear completeness function to degrade the input occurrence, with high completeness to the left (low period) and low completeness on the right (high period). We are able to recover the input occurrence within 1-$\sigma$ of our measurements.}
    \label{fig:toy_gauss}
\end{figure*}

To test the amount of structure we can truly measure, we inject occurrence patterns and see what details we can recover. We do this with four different tests with varying Gaussian occurrence width and separation, shown in the first row of Figure~\ref{fig:toy_gauss}:
\begin{enumerate}
    \item ``small/far'': Gaussian occurrence width of (period width, radius width) = (0.1, 0.1) and separated in radius by log(radius) = 0.5 log \Re{}
    \item ``small/near'': Gaussian occurrence width of (0.1, 0.1) and separated in radius by log(radius) = 0.25 log \Re{}
    \item ``large/far'': Gaussian occurrence width of (0.17, 0.17) and separated in radius by log(radius) = 0.5 log \Re{}
    \item ``large/near'': Gaussian occurrence width of (0.17, 0.17) and separated in radius by log(radius) = 0.25 log \Re{}
\end{enumerate}

For each test, we use a proxy completeness that decreases linearly with increasing orbital period. We use three pairs of Gaussian occurrence for each test to investigate the effect of decreasing completeness and run the injected occurrence through the entire methodology. The second row of Figure~\ref{fig:toy_gauss} shows our measured occurrence. In each of the four tests, the low completeness occurrence at long orbital period is noise-dominated. 

We marginalize each Gaussian occurrence pair and look at their radius distribution, shown in the third row of Figure~\ref{fig:toy_gauss}. We are able to recover the structure (maximum of the two Gaussian occurrence peaks and the minimum between them) for all but one case: the low completeness pair for the large/near case. The measured occurrence is noise-dominated and the measured peak is $0.51\pm0.34$ \Re{} from the input location. We are able to measure structure on the expected scale of the radius valley for completeness values greater than 1\%. 

\begin{figure*}
    \centering
    \includegraphics[width=0.8\textwidth]{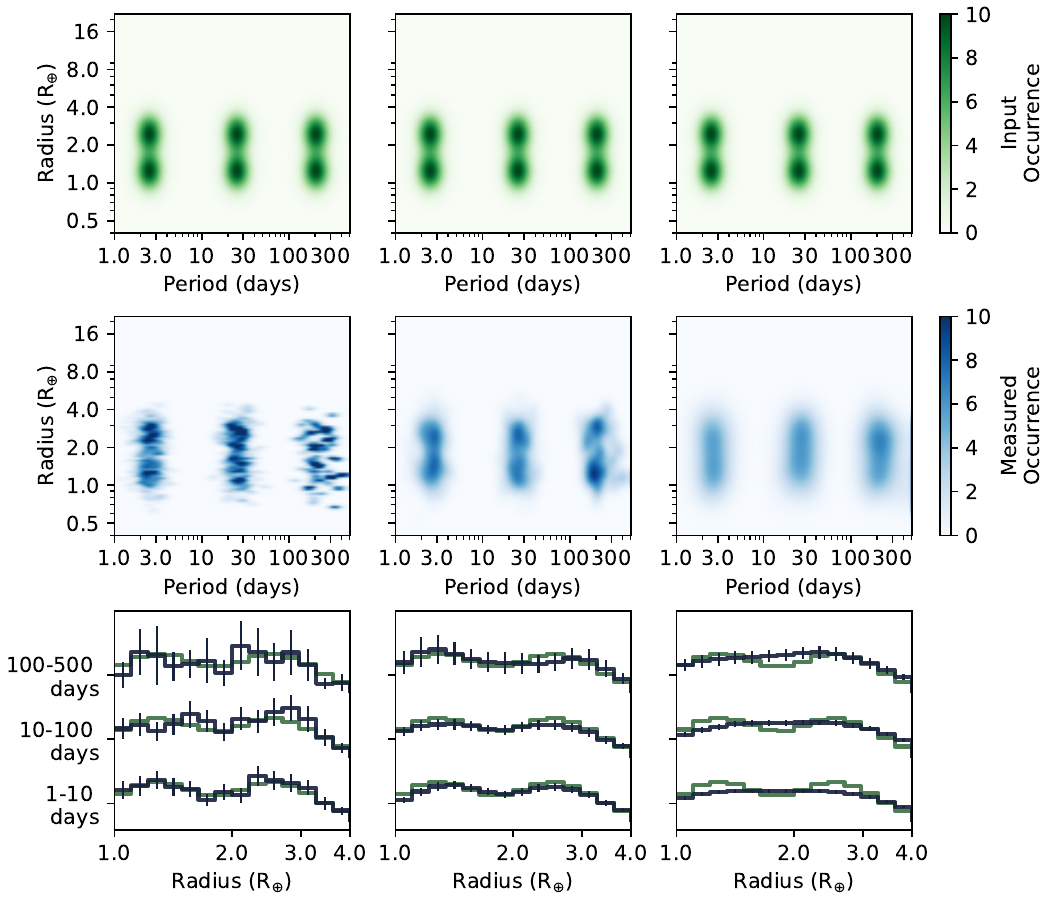}
    \caption{Comparison of different kernel sizes. Each column, from left to right: $(\sigma_x, \sigma_y)=$(0.05, 0.01); (0.05, 0.05); (0.1, 0.1). The top row is the true input occurrence. The middle row is the occurrence returned from our methodology. The bottom row compares the marginalized occurrences with 1-$\sigma$ uncertainties for the measured occurrence. The 1-10 day marginalization for the largest kernel size has uncertainties smaller than the line width.}
    \label{fig:toy_kernels}
\end{figure*}

We empirically determine kernel size by using the small/near injected kernels with three different kernel sizes, shown in Figure~\ref{fig:toy_kernels}. 
The smallest kernel size $(\sigma_x, \sigma_y) = (0.05, 0.01)$ (first column), recovers the injected structure well, with each pair measuring the peak or minimum within 1-$\sigma$ of the true value. However, there is a large amount of fine structure in the marginalized occurrence because the $\sigma_y$ is smaller than our bin size.
The middle column, with $(\sigma_x, \sigma_y) = (0.05, 0.05)$, likewise recovers the input structure within 1-$\sigma$ of the input values for all levels of completeness.
The right column, with $(\sigma_x, \sigma_y) = (0.1, 0.1)$, over-smooths the occurrence distribution and do not measure a valley at all.

For the smallest kernel size, the measured peaks are in agreement with the input occurrence, but the measured distributions are noisy. The largest kernel size over-smooths the distribution, resulting in a lack of any structures. We use a kernel size of $(\sigma_x, \sigma_y) = (0.05, 0.05)$ for the occurrence measurements in Section~\ref{sec:occ}. The choice of kernel does not affect our total occurrence measurement; the total occurrence measured in each case is within 5\% of the input occurrence.  \citet{jin2021} also found this to be true; they used a KDE to measure relative planet occurrence and likewise found that kernel size acted only as a smoothing factor and that overall occurrence was robust to choice of kernel size. The difference in smoothing, however, is significant, which can limit the structure we are able to measure.

\subsubsection{Looking at low completeness and edge effects}\label{sec:uniform}
There are also limits on \textit{where} we are able to measure occurrence with a KDE. We are limited by KDE edge effects and our completeness function.

The KDE is computed for a specific parameter space. Within one kernel-width of the boundary there is a systematic under-reporting of occurrence because the KDE acts as if there is no population on the other side of the boundary as opposed to it being undefined.
Our completeness function also limits where we may measure occurrence: the \Kepler{} completeness function steeply degrades at small radius and long orbital period, creating a region where we expect to detect less than one planet. 

We can test both of these limits by applying a uniform occurrence across the entire parameter space and seeing what we can recover by assuming (1) a uniform, 100\% completeness to test edge effects, or (2) the true \Kepler{} completeness to see where the planet recovery per bin falls to zero. 

\begin{figure*}
    \centering
    \includegraphics[width=\textwidth]{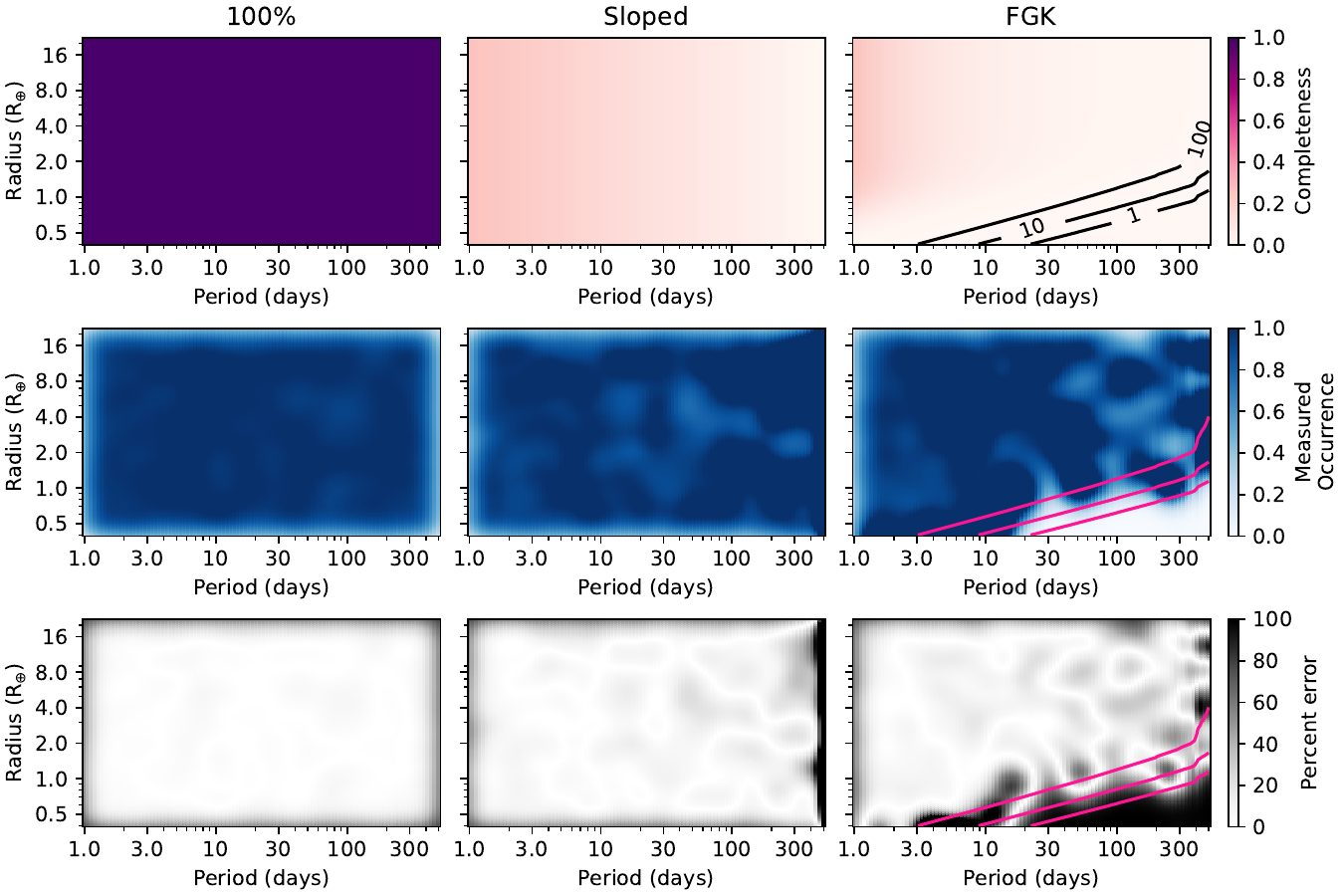}
    \caption{Toy model for three different completeness contours with uniform input occurrence. Left column: 100\% completeness. Middle column: sloped completeness contours used in Figures~\ref{fig:toy_gauss}, \ref{fig:toy_kernels}. Right column: \Kepler{} FGK occurrence. The center row is the measured occurrence from our KDE methodology. The bottom row is the percent error. The leftmost column shows the impacts of edge effects on the resulting KDE. In the rightmost column, we use the middle contour line as our FGK low-completeness boundary.}
    \label{fig:toy_edges}
\end{figure*}

We test this by uniformly populating our parameter space at 10 planets per star per grid cell and using a 100\% completeness function (so there is no degradation of the occurrence). We use the largest tested kernel size of $h=(0.1, 0.1)$ as this will have the largest edge effect. The sample is then bootstrapped 5000 times and averaged for our final measured occurrence. The first column of Figure~\ref{fig:toy_edges} shows the measured occurrence (center panel, in blue) and the residuals (bottom panel, in black) of this test.

\textit{Edge effects:} The density of the KDE will always be lower at the edges; there is no contribution of neighboring Gaussians outside of the specified parameter space. How large of an effect this is depends on the width of the Gaussian kernel. The size of the under-density region is one kernel-width.

We specifically calculate our KDE from $1-500$ days and $0.4-22$\Re{} and only calculate occurrence on the subdomain of $1-400$ days and $0.5-16$\Re{} in order to mitigate these edge effects. This completely removes any edge effects in radius. In period, we only integrate to 400 days in part to avoid the edge effects; however, we are unfortunately limited in our approach at the lower period boundary.

We require occurrence measurements down to 1 day to compare with previous studies (as is the goal of this paper). To avoid edge effects at the low period boundary, we would need to extend our computation domain to 0.9 days for our completeness, reliability, and planet density KDE. 
We do not extend our KDE domain to below 1 day because we do not want to include the spurious detections below one day (discussed in Section~\ref{sec:planets}). Including these planets would extend our domain, therefore avoiding KDE edge effects, but including those spurious planet candidates would impact our occurrence rate measurements more than the edge effects.

KDE edge effects are a limitation of the methodology in general. A different way to mitigate these effects would be to use ``reflected boundaries,'' where it is assumed that the KDE is constant on the other side of the boundary \citep{silverman1986}. We do not believe this would be an appropriate assumption for low period planets due to the small number of planet detections in this regime.

Further, the region where these edge effects are impactful is only from $1-1.12$ days for our kernel size. These edge effects are the most extreme when there is a high KDE density right at the boundary. For low orbital period, between $1-1.12$ days, the low density affects the occurrence measurement in Figure~\ref{fig:toy_edges} by about 27\%. $1-1.12$ days is small relative to the bin size we use throughout the paper. How much this affects the reported occurrence values scales with the bin width; for the binning scheme presented in Figure~\ref{fig:FGK}, this amounts to $\sim4\%$ of the reported occurrence for the lowest period bin, $1.0-2.11$ days.

\textit{Low completeness:} The low-completeness limit can also be investigated by setting a uniform occurrence and using the real FGK completeness function. 

We measure this in the third column of Figure~\ref{fig:toy_edges}. The three contours represent the completeness level of 100/$N_{stars}$, 10/$N_{stars}$, and 1/$N_{stars}$ where $N_{stars}$ is the total number of stars in our sample. A conservative threshold of low completeness can be set at the 100/$N_{stars}$ level; there is complete recovery above this contour. While not complete recovery at the 10/$N_{stars}$ level, there is still some recovery and we set this as our low-completeness boundary. Below this contour, there is very little recovery, and none below the 1/$N_{stars}$ level. We only report upper limits on the occurrence below the 10/$N_{stars}$ level.

\section{Occurrence}\label{sec:occ}

\begin{figure*}
    \centering
    \includegraphics[width=\textwidth]{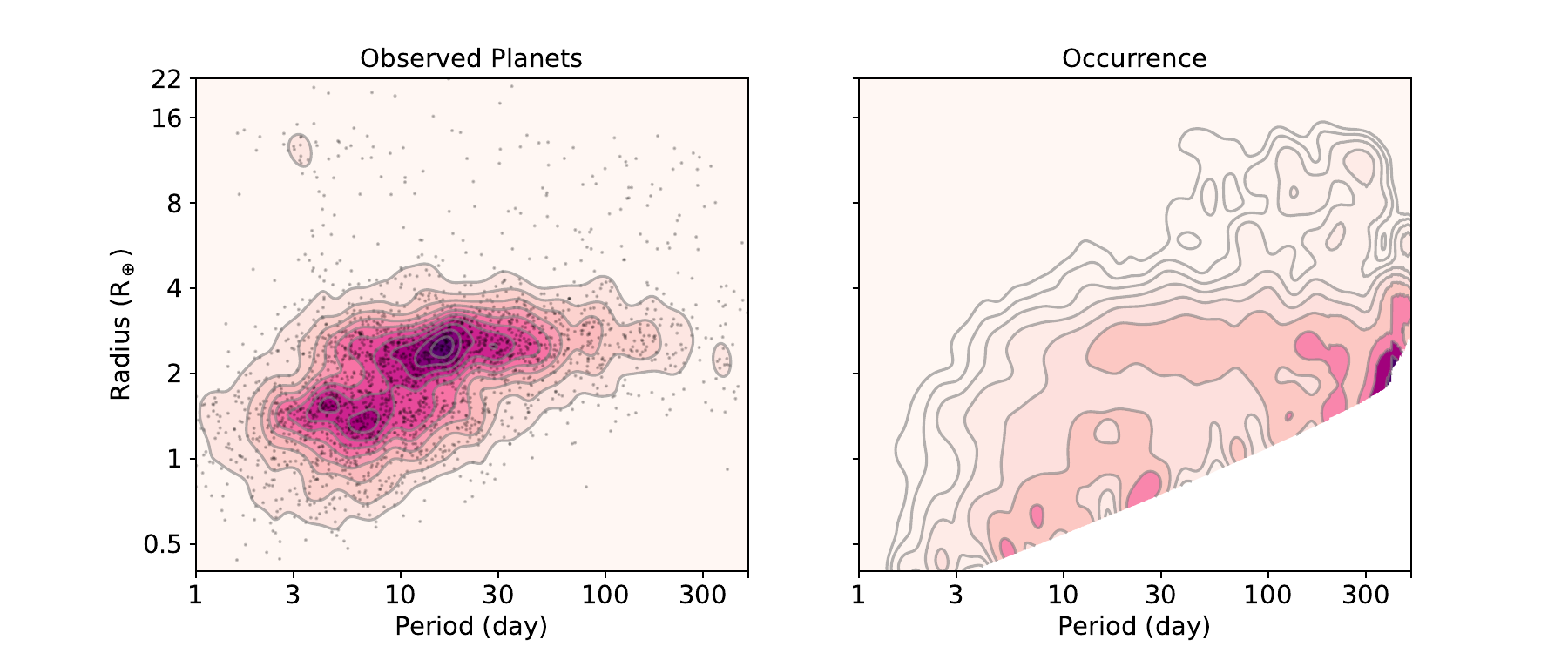}
    \caption{KDE maps of the observed FGK planet population (left) and the completeness-corrected intrinsic occurrence (right).}
    \label{fig:kde}
\end{figure*}

We have described a non-parametric technique for computing occurrence rates. The long-term goal of this methodology is to map features in multidimensional parameter space to theoretical models of planet formation and evolution. The observed planet KDE and its occurrence counterpart can be seen in Figure~\ref{fig:kde}. There is a wealth of features to be seen in the occurrence map. There appears to be a limit in period to the small planet distribution, tracing out the lower boundary to the Neptune desert. The radius valley can clearly be seen just below 2 \Re{}. Above the large population of sub-Neptunes, there is a decline in occurrence. Further, the shape of this drop-off in occurrence is not constant across different orbital periods. Not all of these structures can be well-described with analytic functions, so a non-parametric approach is key to be able to describe these structures. In order to tie these structures in with theory, we must first validate our methodology by comparing benchmark measurements against literature values. To do this, we bin the parameter space, integrate occurrence over specific domains, and collapse the 2D map into 1D distributions for comparisons. In Section~\ref{sec:integrated} we compute integrated rates; in Section~\ref{sec:rad} we look at the 1D radius distribution; in Section~\ref{sec:period} we look at the 1D period distribution; in Section~\ref{sec:type} we look at these as a function of stellar type. In Section~\ref{sec:cliff} we explore the occurrence cliff.

\subsection{Integrated Rates}\label{sec:integrated}

\begin{figure*}
    \centering
    \includegraphics[width=\textwidth]{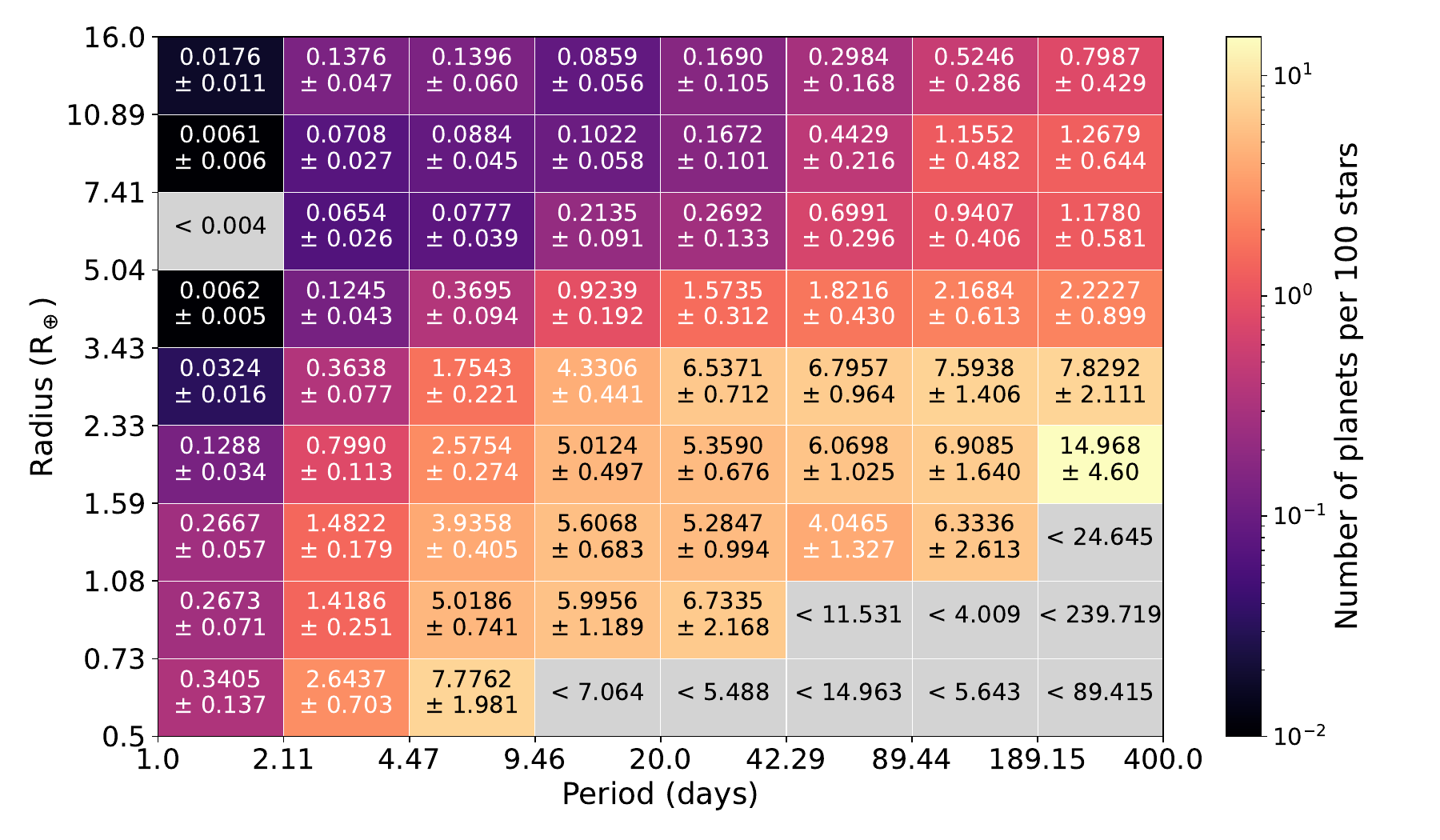} 
    \caption{FGK occurrence rate estimates. The number of planets per 100 stars (shown) is comparable to a percentage. Gray cells are those with either no planet detections or below the low occurrence boundary (see Section~\ref{sec:completness}), with only the upper limit shown.}
    \label{fig:FGK}
\end{figure*}

Figure~\ref{fig:FGK} shows our occurrence rate estimates and 1-$\sigma$ uncertainties for the entire period and radius range of $1-400$ days and $0.5-16$ \Re{}. Gray bins either represent bins with no planet detections in our chosen binning scheme or bins that fall below the completeness boundary established in Section~\ref{sec:toy}; they contain upper limits for the occurrence. We use the log-uniform bins $P=$[1.0, 2.11, 4.47, 9.46, 20.0, 42.29, 89.44, 189.15, 400.0] days in period and $R_P=$[0.5, 0.73, 1.08, 1.59, 2.33, 3.43, 5.04, 7.41, 10.89, 16.0] \Re{} in radius to minimize the number of empty bins in our sample.

Integrating over bins with detections above the low completeness boundary (colored bins in Figure~\ref{fig:FGK}), we measure $1.52\pm0.08$ planets per star. \citet[][hereafter P18]{petigura2018} report an average value of 1.10 NPPS (albeit for a smaller parameter space; from Figure~6 of their paper), \citetalias{kunimoto2020b} report $1.06^{+0.09}_{-0.07}$ NPPS, and \citetalias{hsu2019} measure a lower limit of $2.37^{+0.93}_{-0.67}$ NPPS (for periods out to 500 days; derived from Figure~2 of their paper) on total FGK occurrence.

While the differences between these measurements are significant, they are not comparable quantities.
The integrated occurrence rate does not serve as a reliable benchmark because integrating over such a large domain encompasses regions of parameter space with zero detections and/or exceptionally low completeness. The degree to which this impacts integrated occurrence depends not only on the input assumptions but also on the details of sample selection, which varies between studies.
The benchmarks discussed in subsequent sections, however, are based on marginalized distributions that better facilitate comparisons.

A more reliable benchmark is the hot Jupiter domain, where completeness is high. The \Kepler{} hot Jupiter occurrence rate has stayed almost constant over the past decade. Between $8-16 R_{\oplus}$ and over a period of $1-10$ days we measure an occurrence rate of $\eta = 0.46\pm0.1\%$. Our measurement is in agreement with the literature values, which range from $0.4-0.6\%$ \citep{howard2012, fressin2013, mulders2015a, petigura2018, kunimoto2020b}. However, this rate is roughly half of the rate measured from radial velocity surveys, which measure between $0.8-1.2\%$ \citep{mayor2011, wright2012, wittenmeyer2020}. 
The pile-up could be a result of metal-rich stars, while the \Kepler{} sample has systematically lower-metallicity stars than RV samples \citet{wright2012, dawson2013}. Studies on the dependence of giant planet occurrence on host star metallicity have found a strong correlation between the two \citepalias{petigura2018}. However, this is still debated as \citet{guo2017} found that metallicity alone cannot explain the difference in rates. It could possibly be explained by differences in false positive rates \citep{wang2015, guo2017}.
The \Kepler{} results have been shown to be within 1-$\sigma$ of the large uncertainty associated with the RV data \citep{fernades2019}.
Recently, \citet{beleznay2022} measured the hot Jupiter occurrence rate for AFG stars with \textit{TESS} and found a rate of $0.39\pm0.06\%$ for F and G stars, which are within 1-$\sigma$ of our result.

\subsection{The Small Planet Radius Distribution}\label{sec:rad}
The radius distribution of small planets ($<4 R_{\oplus}$) is rife with structure. The large population of observed planets can be split into the dense super-Earths and the puffier sub-Neptunes, separated by a region of low occurrence called the radius valley. The bimodal population was first observationally seen with the California-Kepler Survey \citep[CKS;][]{petigura2017} in \citet{fulton2017}, following years of theoretical predictions \citep[e.g.][]{lopez2013, Lee2016, owen2017, gupta2020, rogers2021}. The exact location and slope of the radius valley have been the focus of many studies, as it could inform which physical process is the primary driver of planet formation (see Table 6 of \citet{ho2023} for a summary of values).

\begin{figure}
    \centering
    \includegraphics[width=\columnwidth]{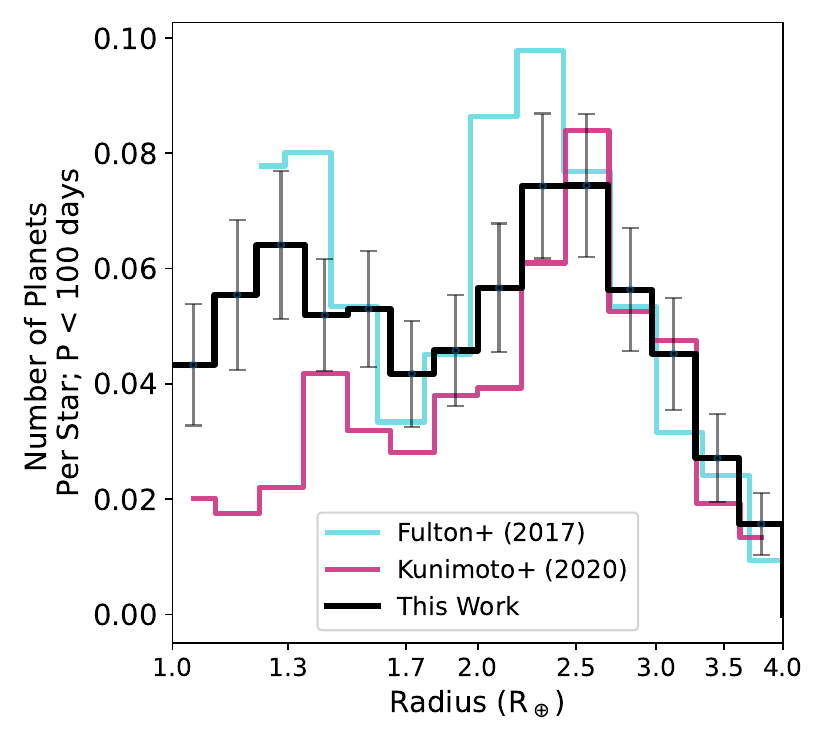}
    \caption{Radius Distribution for small planet with $P < 100$ days. Blue is data taken from \citet{fulton2017} and pink is data taken from \citetalias{kunimoto2020b}. We find the minimum of the radius valley to be $1.78^{+0.14}_{-0.16}$ \Re{}.}
    \label{fig:radius_gap}
\end{figure}

We compare the radius distribution within 100 days with selected previous studies in Figure~\ref{fig:radius_gap}. Only considering occurrence within 100 days not only agrees with the limit to most atmospheric mass loss processes but allows us to integrate over a range where we are not hindered by low completeness (see Section~\ref{sec:toy} for bounds). \citet{fulton2017} (blue in fig.~\ref{fig:radius_gap}) measured this structure with the CKS sample, a precisely-characterized sub-sample of the \Kepler{} dataset, including 36,075 stars and 900 planets. \citet{fulton2017} used an independent planet detection pipeline and associated injection/recovery experiment to correct for completeness; we compare it here to benchmark the evolution of radius valley measurements. \citetalias{kunimoto2020b} (pink in fig.~\ref{fig:radius_gap}) used their own independent pipeline to detect and vet planets and calculated their own reliability and completeness. They incorporate \textit{Gaia} DR2-derived stellar radii to update planet radii. Their catalog of 96,280 FGK stars and 2829 planets is similar to ours but not the same. Overall, we find good agreement for all three distributions. We compare specific features of the distribution next.

There are three distinct structures we measure in the radius distribution: the minimum of the radius valley, the peak of the super-Earths (which we define as the portion of the distribution to the left of the valley), and the peak of the sub-Neptunes (which we define as the portion of the distribution to the right of the valley). We measure them as follows: for each iteration of our bootstrap simulation, we integrate the occurrence within 100 days binned to 50 log-uniform bins between $1-4$ \Re{}. Then we find the minimum occurrence bin between $1.4-2.2$ \Re{} and set that radius, $R_{min}$, as the minimum of the valley. Finally, we re-bin to the final binning scheme ($R_P = $[1.0, 1.1, 1.21, 1.35, 1.49, 1.64, 1.81, 2.0, 2.21, 2.44, 2.69, 2.97, 3.28, 362, 4.0] \Re{}); we use more bins than in the 2D occurrence maps both to resolve the radius valley and to compare with \citetalias{kunimoto2020b}. We fit both planet populations with a Gaussian distribution, from [1.0, $R_{min}$] to the left and [$R_{min}$, 4.0] to the right with \texttt{scipy curve\_fit}, giving us the location of the peaks of the sub-Neptunes and the super-Earths. 
For our FGK population, we find the peak of the sub-Neptunes to be at $2.51^{+0.06}_{-0.08}$ \Re{}, the minimum of the radius valley to be at $1.78^{+0.14}_{-0.16}$ \Re{}, and the super-Earth peak to be at $1.37^{+0.19}_{-0.44}$ \Re{}. 
These measurements are within 1-$\sigma$ agreement to the peak and minimum measurements reported in previous studies. \citet{fulton2017} finds the super-Earth peak to be 1.3 \Re{} and the sub-Neptune peak to be 2.4 \Re{} while \citetalias{kunimoto2020b} finds the super-Earth peak to be 1.4 \Re{}, the minimum of the valley to be 1.7 \Re{}, and the sub-Neptune peak to be 2.6 \Re{}.
\citet{martinez2019} derive precise, spectroscopically-derived radii of 1633 planets and obtain radius distribution measurements of $1.47\pm0.05$ \Re{}, $1.89\pm0.07$ \Re{}, and $2.72\pm0.10$\Re{}, which are in agreement with our measurements within 1-$\sigma$.

It is not obvious that the super-Earth occurrence below the radius valley turns over toward smaller radii (that there is a true super-Earth ``peak,'' in other words). While \citet{fulton2017} reported a bimodal distribution with two distinct peaks, \citetalias{kunimoto2020b} found a flattening of the population below the valley, and \citetalias{hsu2019} found an increasing occurrence of planets below the valley down to 0.5 \Re{}.
We fit the super-Earths ($P=$[1, 100] days and $R=$[1, 1.78] \Re{}) with both a Gaussian distribution and linear function to differentiate between these scenarios. To tell which model is better, we compute the Akaike Information Criterion (AIC) \citep{akaike1974}. The (minimum, mean) AIC for the Gaussian fit is (-33.04, -26.22) while the (minimum, mean) AIC linear fit is (-30.48, -27.59). The difference in AIC minimums is less than 3, which constitutes no substantial evidence that either model is preferred, so we also compare their Bayesian Information Criterion (BIC) \citep{schwarz1978}. The (minimum, mean) BIC for the Gaussian fit is (-36.85, -30.03) while the (minimum, mean) BIC for the linear fit is (-33.09, -30.27). $\Delta BIC_{min} = 3.76$, which means that the Gaussian fit is slightly favored but still does not constitute substantial evidence of a preferred model. We do not have precise enough occurrence measurements below the valley to say if there is a super-Earth peak or if occurrence flattens out below the valley. 

While we find no substantive evidence that the population within 100 days turns over at 1 \Re{}, we can look at the population below 1 \Re{} to see if there is a peak in occurrence for the terrestrials, which we define here as the planets with R$=[0.7-1.0]$ \Re{}. We limit our orbital period range to 30 days for this analysis, in order to stay within our low completeness boundary. Overall, occurrence is higher at small radius, with large error bars. We compute the same fits as before, a Gaussian distribution and a linear fit, for the smaller orbital period range ($P=$[1, 30] days and $R=$[0.7, 1.75] \Re{}). $\Delta AIC_{min} = 1.85$ and $\Delta BIC_{min} = 0.66$ when comparing the Gaussian and linear fits, so we are unable to say which model is preferred. We find less evidence of a peak in occurrence for the terrestrials than for the super-Earths. Substantial uncertainties on occurrence ($\sim23\%$ relative errors compared to $<20\%$ for the 100 day occurrence) are likely the driver of the uncertainty in the fits. In comparison, \citetalias{hsu2019} measure occurrence down to 0.5 days and found increasing occurrence down to small radii. Even below 30 days, we do not measure the substantial population of terrestrials that \citetalias{hsu2019} does; our occurrence for the bin P=[1., 32.] days and $R_P$=[0.75, 1.] \Re{} is $F=0.139\pm0.03$ where theirs is $F=0.186\pm0.04$.

\subsection{Period Distribution}\label{sec:period}
We marginalize our FGK occurrence rates into four different radius bins to investigate the period distribution. The four radius bins are $1-2$ \Re{} (which we refer to as super-Earths in this section), $2-4$ \Re{} (which we refer to as sub-Neptunes in this section), $4-8$ \Re{}, and $8-16$ \Re{}. We use these specific radius bins to directly compare with previous studies, instead of using the radius valley minimum measured in Section~\ref{sec:rad} as the dividing radius between super-Earths and sub-Neptunes.

\begin{figure}
    \centering
    \includegraphics[width=\columnwidth]{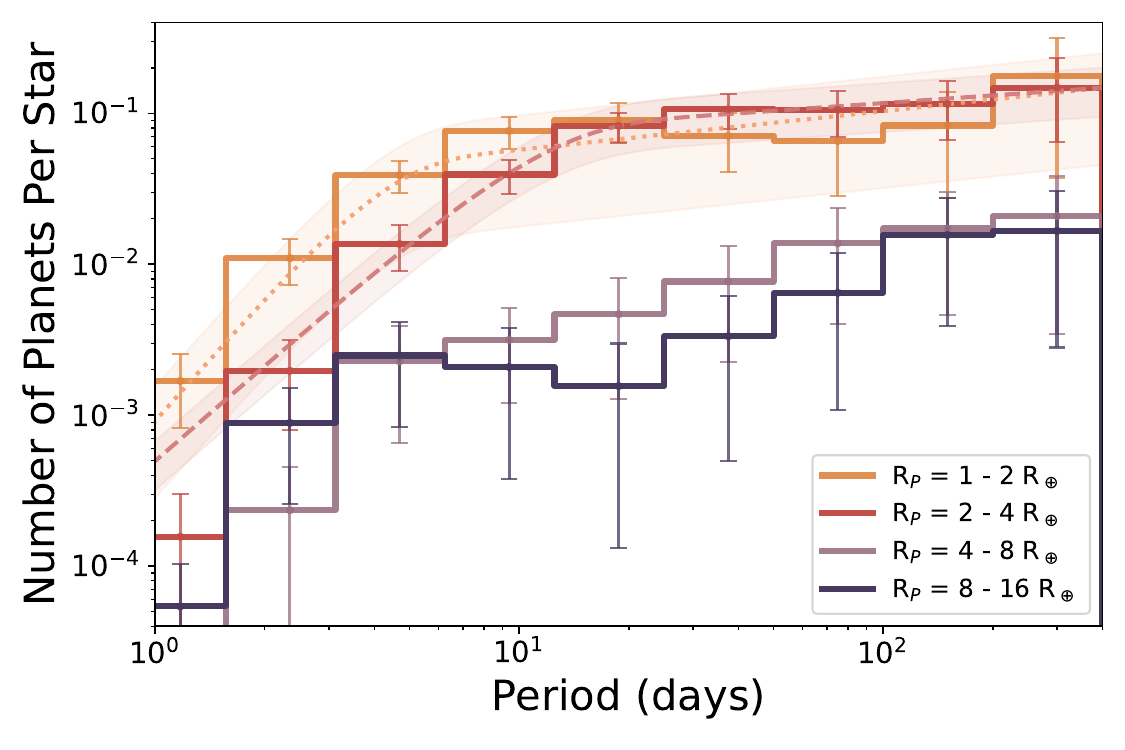}
    \caption{The period distribution of small planets, with fits to Eq.~\ref{eq:period} for the small planets.
    The terrestrial planets, $1-2 R_{\oplus}$, are plotted in orange and the sub-Neptune planets, $2-4 R_{\oplus}$ are plotted in pink. Both small planet populations show a rise in occurrence until a break period, where occurrence flattens out. The larger planets, in bins $4-8 R_{\oplus}$ (purple) and $8-16 R_{\oplus}$ (navy), show no break point and rise in occurrence across all orbital periods.}
   \label{fig:period_dist}
\end{figure}

The period distribution can be seen in Figure~\ref{fig:period_dist}\footnote{Figure~\ref{fig:period_dist} shows the integrated occurrence within each histogram bin, $F$, or Number of Planets Per Star, to be consistent with all other plots in the paper. Equation~\ref{eq:period} is given for the occurrence \textit{rate}, or $\frac{df}{d\log{P}}$. To convert between the two, we multiply by the bin size, or $\log{P}=0.25$. Comparisons can be made to Figure 7 of \citetalias{petigura2018}, which shows the period distribution as a rate, while Figure 6 of \citet{howard2012} and Figure 12 of \citetalias{kunimoto2020b} show the integrated rates, as we do here.}. Overall, the small planets, $1-2$ \Re{} and $2-4$ \Re{}, show a sharp increase in occurrence until a break period, where occurrence flattens out. The large planets do not show evidence of a period break and slowly rise in occurrence with increasing orbital periods.

We fit the small plant distributions with a broken power law as in \citet{howard2012}:
\begin{equation}\label{eq:period}
    \frac{df}{d\log{P}} = C P^\beta (1 - \exp^{(-P/P_0)^\gamma})
\end{equation}

We fit this distribution with the \texttt{scipy curve\_fit} package, via the same method as the radius distribution fit. The broken power law simplifies into a piecewise function of the form:

\begin{equation}
    \frac{df}{d\log{P}} \propto \begin{cases}
        P^{\alpha} & \text{if $P \ll P_0$, where $\alpha = \beta + \gamma$}\\
        P^{\beta} & \text{if $P \gg P_0$}
    \end{cases}
\end{equation}

The steep rise in occurrence, with slope $\alpha$, traces out the edge of the Neptune desert at short orbital period. The sharp rise in occurrence implies there is a limit to how close planets can reside to their host star. Processes that result in this limit could be the disk inner edge inhibiting planet formation, the disk edge trapping migrating planets, or tidal interactions removing planets. Past a break point $P_0$, the occurrence flattens out. The specific period $P_0$ where this transition occurs could also relate to limits of where planetary migration or formation is. Differences between the $1-2$ \Re{} and $2-4$ \Re{} break locations can indicate how those formation processes depend on planetary size.

All of the fit parameters and 68.3\% confidence intervals for super-Earths ($1-2$ \Re{}) and sub-Neptunes ($2-4$ \Re{}) can be found in Table~\ref{tab:period}. For planets $1-2$ \Re{}, occurrence rises with a slope of $\alpha=2.76^{+0.29}_{-0.20}$  to a transition period of $P_0=4.10^{+0.58}_{-0.66}$ days. The slope is much steeper, and transition period much smaller than those measured in \citetalias{kunimoto2020b} ($\alpha=1.9^{+0.1}_{-0.1}, P_0=5.9^{+0.5}_{-0.5}$) or \citetalias{petigura2018} ($\alpha=2.4^{+0.4}_{-0.3}, P_0=6.5^{+1.6}_{-1.2}$). The difference might be explained by \citetalias{petigura2018} measuring only $1-1.7$ \Re{} for their fits. We notably measure a much higher occurrence of super-Earths than \citetalias{kunimoto2020b} (as in the radius distribution in Figure~\ref{fig:radius_gap}), which appears here as a steeper slope and earlier period break.

For the sub-Neptunes, $2-4$ \Re{}, we measure $\alpha=2.11^{+0.18}_{-0.17}$ and $P_0=11.12^{+1.14}_{-0.93}$. These are consistent with $\alpha=2.2^{+0.1}_{-0.1}$ and $P_0=13.3^{+1.4}_{-1.5}$ from \citetalias{kunimoto2020b} and $\alpha=2.3^{+0.2}_{-0.2}$ and $P_0=11.1^{+1.7}_{-1.5}$ from \citetalias{petigura2018}.

We measure an increase in occurrence out to long orbital period for both super-Earths and sub-Neptunes, with $\beta = 0.26^{+0.09}_{-0.08}$ for $1-2$ \Re{} planets. A positive slope is in conflict with previous studies that measure a decrease in the occurrence of small planets: \citetalias{kunimoto2020b} measures the slope to be decreasing with $\beta = -0.5^{+0.1}_{-0.1}$ for $1-2$ \Re{} and \citetalias{petigura2018} measures $\beta = -0.3^{+0.2}_{-0.2}$ for $1-1.7$ \Re{}. Even \citet{dong2013}, which measures a relatively flat slope at $\beta=-0.10\pm0.12$ for $1-2$ \Re{}, is outside of our uncertainties.

For $2-4$ \Re{} planets, we also measure a slope higher than previous studies, with $\beta = 0.17^{+0.04}_{-0.05}$. Both \citetalias{kunimoto2020b} and \citetalias{petigura2018} measure a nearly flat slope of $\beta = -0.1^{+0.1}_{-0.1}$. \citet{dong2013} does measure a rise in slope for this radius range, but it is still outside of our uncertainties at $\beta = 0.11\pm0.05$.

We attribute this rise in slope to the higher occurrence of planets we find in the last period bin, $189.15-400$ days. 
For both radius ranges, this period bin is the main driver of the positive $\beta$ slope. \citetalias{petigura2018} does not measure past 300 days and so does not see this increase. \citetalias{kunimoto2020b} systematically measures lower planet occurrence than our measurements below 2 \Re{}. \citetalias{hsu2019} does not fit any distributions to their occurrence rate estimates, but they do measure an increased occurrence for their last period bin, 256-500 days.
If we exclude the bin, we measure slopes of $\beta = -0.086^{+0.01}_{-0.01}$ for the $1-2$ \Re{} bin and $\beta = 0.089^{+0.02}_{-0.01}$ for the $2-3$ \Re{} bin. Without the last bin, super-Earth slope is negative and in line with \citet{dong2013} and \citetalias{petigura2018}. The sub-Neptune slope is still positive, but without the last bin is comparable to \citet{dong2013}.
Further, if we were to take into account the effects of transit multiplicity into account, we would see an even more increased positive slope relative to the other measurements. 

The larger planet populations, in the $4-8 R_{\oplus}$ and $8-16 R_{\oplus}$ bins, do not show a break in occurrence but are instead a steady rise across all orbital periods. We do not see evidence for a three-day pile-up in the hot Jupiters, as found in radial velocity surveys \citep{mayor2011, wright2012, weiss2014, wittenmeyer2020}. Studies from \Kepler{} have likewise not seen this overabundance \citep{howard2012, fressin2013, kunimoto2020b}.

\begin{deluxetable*}{ccccc}

\tabletypesize{\footnotesize}
\tablecolumns{5}
\tablewidth{0pt}
\tablecaption{ Period Distribution Fit \label{tab:period}}

\tablehead{
    \colhead{$R_P$ (\Re{})} & \colhead{C} & \colhead{$\beta$} & \colhead{$\gamma$} & \colhead{$P_0$ (days)}} 
    
\startdata \\[-3mm]
        1 - 2 & $ 0.13^{+0.05}_{-0.03}$ & $ 0.26^{+0.09}_{-0.08}$ & $ 2.5^{+0.20}_{-0.12}$ & $ 4.10^{+0.58}_{-0.66}$ \\[2mm]
        2 - 4 & $0.21^{+0.5}_{-0.04}$ & $ 0.17^{+0.04}_{-0.05}$ & $ 1.94^{+0.14}_{-0.12}$ & $ 11.12^{+1.14}_{-0.93}$ \\[2mm]
\enddata

\tablecomments{Median and 68.3\% confidence intervals for the fits to the parametric period distribution from Eq.~\ref{eq:period}.}

\end{deluxetable*}

\subsection{Stellar Type Dependence}\label{sec:type}
\begin{figure*}
    \centering
    \includegraphics[width=\textwidth]{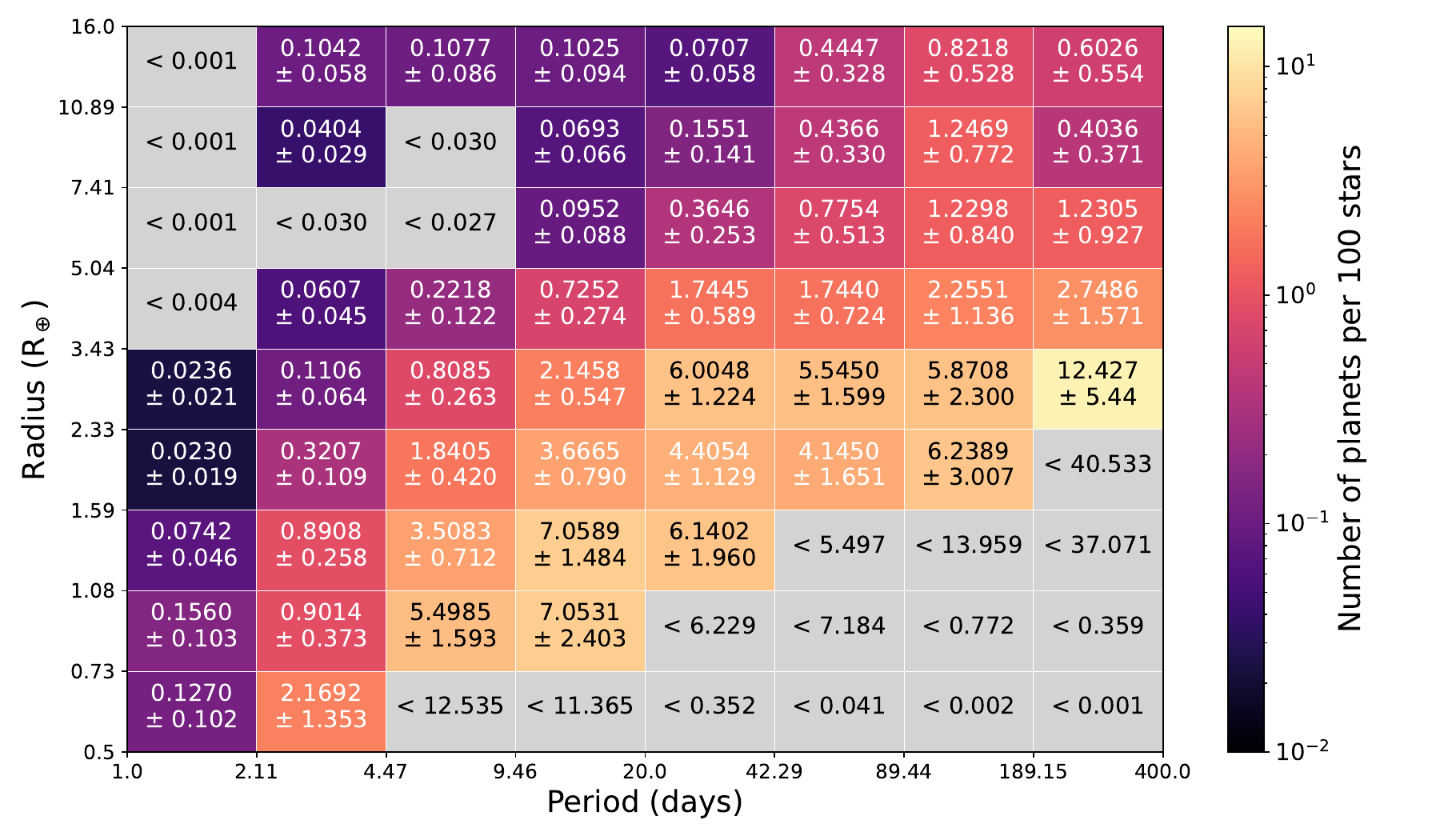}
    \caption{Same as Figure~\ref{fig:FGK}, calculated only for F-type stars}
    \label{fig:F}
\end{figure*}

\begin{figure*}
    \centering
    \includegraphics[width=\textwidth]{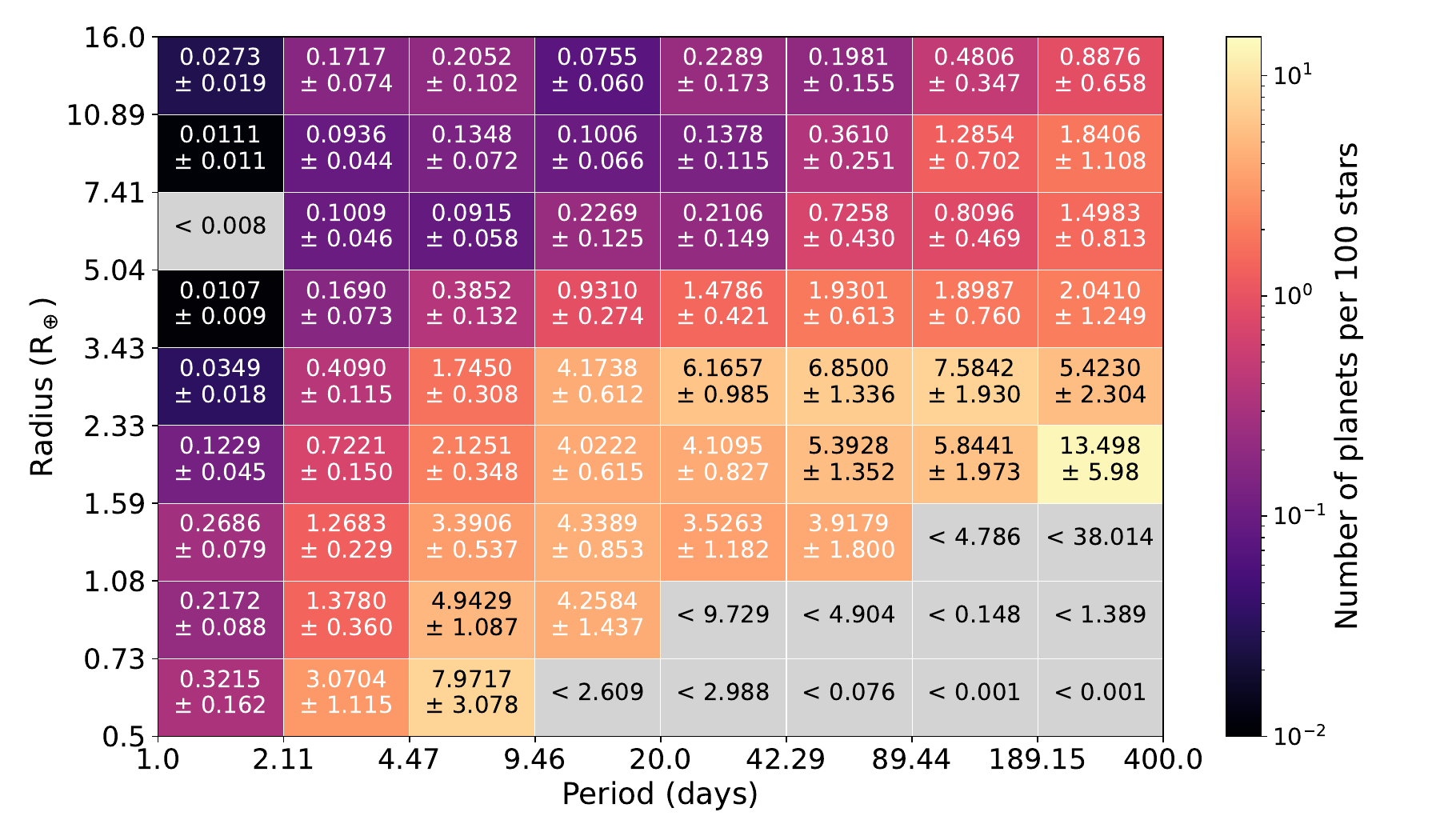}
    \caption{Same as Figure~\ref{fig:FGK}, calculated only for G-type stars}
    \label{fig:G}
\end{figure*}

\begin{figure*}
    \centering
    \includegraphics[width=\textwidth]{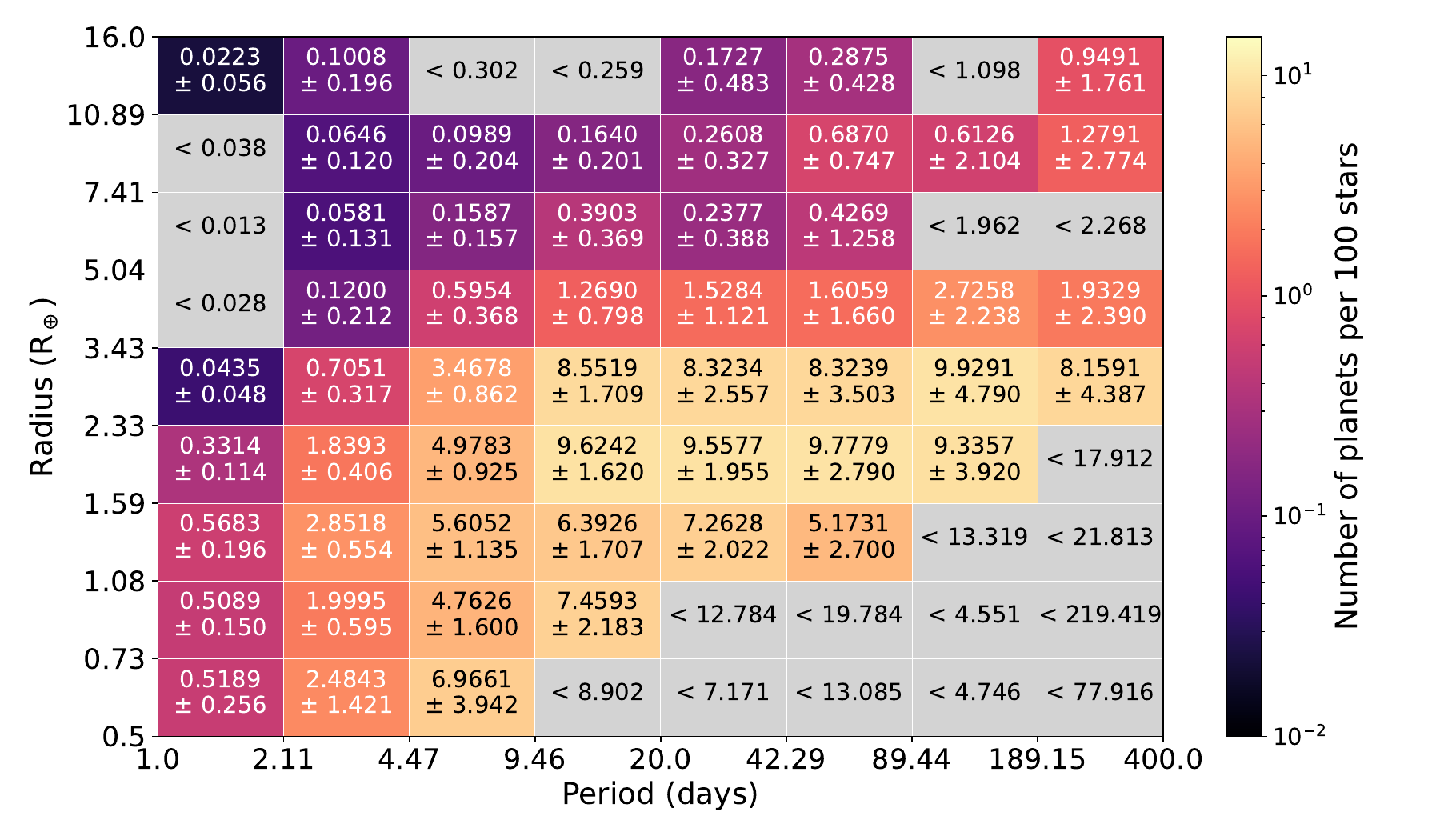}
    \caption{Same as Figure~\ref{fig:FGK}, calculated only for K-type stars}
    \label{fig:K}
\end{figure*}

We separate the parent FGK population into F, G, and K stars based on the effective temperature boundaries from \citet{pecaut2013}. Figure~\ref{fig:F} shows the occurrence for F stars $(6000$ K $\leq T_{eff} < 7300$ K), Figure~\ref{fig:G} for G stars $(5300$ K $\leq T_{eff} < 6000$ K), and Figure~\ref{fig:K} for K stars $(3900$ K $\leq T_{eff} < 5300$ K).

The occurrence between $1-400$ days and $0.5-16$ \Re{} is $1.05\pm0.09$ planets per F star; $1.26\pm0.09$ planets per G star; and $1.61\pm0.13$ planets per K star. As in Section~\ref{sec:integrated}, these numbers are lower limits to occurrence and are the sum of filled cells in their respective figures; if combined and weighted appropriately, they do not add up to the given FGK occurrence. The total occurrence calculated above the strictest occurrence boundary but including the upper limits for otherwise empty cells (e.g., the Neptune desert) is $1.01 \pm 0.09$ planets per F star, $0.97 \pm 0.06$ planets per G star, and $1.47 \pm 0.12$ planets per K star\footnote{Following, the total FGK occurrence calculated above the strictest completeness boundary (the K star boundary) but including the upper limits for otherwise empty cells is $1.09\pm0.04$. These measurements are then within firm agreement with the previously cited studies.}.
We measure a significant increase in planet occurrence toward cooler stars.

The relationship between increased planet occurrence at lower stellar effective temperature has been well documented in occurrence rate studies \citep{howard2012, mulders2015c, kunimoto2020b, yang2020}.
In each study, an increased occurrence for later-type stars is measured.
The physical cause of the increased occurrence is not well understood. \citet{mulders2015c} found that the increasing occurrence rates for later stellar types (M dwarf compared to FGK) cannot be simply explained by redistributing the same amount of heavy element mass of larger (sub-Neptune-sized) planets in systems around earlier-type systems into many smaller planets.

\citetalias{kunimoto2020b} is the closest comparable study to ours. They report a much stronger trend of increasing occurrence with cooler stars: they report $0.89^{+0.23}_{-0.16}$ planets per F star; $1.67^{+0.21}_{-0.16}$ planets per G star; and $2.56^{+0.29}_{-0.24}$ planets per K star. These differences in integrated occurrence are subject to the same issues that prevent quantitative comparison as discussed in Section~\ref{sec:integrated}.

\subsubsection{Radius Valley as a function of stellar type}
We measure the features of the radius distribution via the same method as we did for the FGK sample for each stellar type. This analysis supposes there is an underlying two-peak structure to the occurrence. Because F, G, and K total occurrence differ, this structure is not as obvious by-eye as it is for the FGK occurrence. The choice of \textit{two} peaks may also be suspect, as discussed at the end of Section~\ref{sec:rad}.

\begin{figure}
    \centering
    \includegraphics[width=\columnwidth]{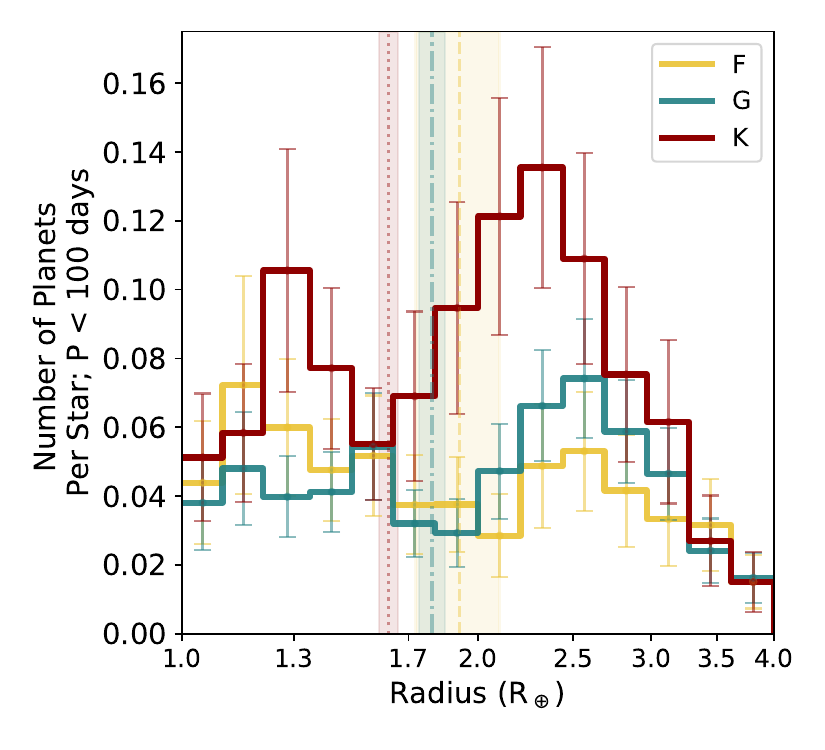}
    \caption{Population of small planets within 100 days (same as in Figure~\ref{fig:radius_gap}, now with valley minimums marked with vertical lines), now separated by stellar type. The minimum of the radius valley moves to smaller radius for cooler stars: $R_{min}=1.92^{+0.26}_{-0.18}$ \Re{} for F stars (red; dotted),$R_{min}=1.80^{+0.16}_{-0.05}$ \Re{} for G stars (teal; dot-dashed), and $R_{min}=1.62^{+0.02}_{-0.04}$ \Re{} for K stars (yellow; dashed), compared to $R_{min}=1.78^{+0.14}_{-0.16}$ \Re{} for all FGK stars.}
    \label{fig:stellar_rg}
\end{figure}

As seen in Figure~\ref{fig:stellar_rg}, there is a substantial shift in the minimum of the valley between stellar types: the minimum of the valley shifts to smaller radii with lower stellar effective temperature. We measure the minimum of the valley to be $R_{min}=1.92^{+0.26}_{-0.18}$ \Re{} for F stars, $R_{min}=1.80^{+0.16}_{-0.05}$ \Re{} for G stars, and $R_{min}=1.62^{+0.02}_{-0.04}$ \Re{} for K stars. As with the FGK sample, we measure the location of the super-Earth and sub-Neptune peaks individually for each stellar type. The sub-Neptune peak follows the minimum of the radius valley and moves towards smaller radii for cooler stellar types. Individual measurements are reported in Table~\ref{tab:radius}.

\begin{deluxetable*}{c|rrrr}
\tabletypesize{\footnotesize}
\tablecolumns{5}
\tablewidth{0pt}
\tablecaption{ Radius Valley Fits \label{tab:radius}}

\tablehead{\colhead{} & \colhead{SE Peak} & \colhead{Minimum} & \colhead{SN Peak} & \colhead{Ratio}}
\startdata \\[-3mm]
    FGK & 
    $1.37^{+0.19}_{-0.44}$ & $1.78^{+0.14}_{-0.16}$ & $2.51^{+0.06}_{-0.08}$ &$2.24^{+0.78}_{-0.64}$ \\[2mm]
    F & 
    $1.28^{+0.32}_{-0.76}$ & $1.92^{+0.26}_{-0.18}$ & $2.58^{+0.19}_{-0.59}$ &$1.37^{+0.83}_{-0.52}$ \\[2mm]
    G & 
    $1.41^{+0.24}_{-0.59}$ & $1.80^{+0.16}_{-0.05}$ & $2.61^{+0.07}_{-0.08}$ &$2.50^{+0.64}_{-0.71}$ \\[2mm]
    K & 
    $1.39^{+0.25}_{-0.12}$ & $1.62^{+0.02}_{-0.04}$ & $2.34^{+0.08}_{-0.09}$ & $3.41^{+0.77}_{-0.68}$ \\[2mm]
\enddata
\tablecomments{Measured locations of radius distribution features for different stellar types. Each location given in \Re{}.}
\end{deluxetable*}

We can also compare the stellar type dependence of the radius valley to previous studies if we take stellar effective temperature as a proxy for stellar mass. For main sequence stars, mass and effective temperature correlate. We can derive the relationship via the Stefan-Boltzmann law and the classical mass-luminosity relationship $L = 4\pi R^2 \sigma T_{eff}^4$ and $\frac{L}{L_\odot} = \frac{M}{M_\odot}^{\alpha_C}$ where classically $\alpha_C = 3.5$. \citet{eker2018} uses a variable mass-luminosity relationship for different mass ranges, with $\alpha_C$ ranging from $2.028-5.743$. For all proposed values of $\alpha_C$, the relationship has a positive correlation and classically is $M \propto T^\frac{8}{7}$.

The stellar mass dependence of the radius valley has been investigated in \citet{wu2019}, \citet{berger2020a}, \citet{petigura2022}, and \citet{ho2023} by quantifying the slope, $\alpha$, of the valley in the stellar mass versus planet radius plane, with values ranging from $\alpha=0.23-0.35$. 
\citet{wu2019} investigated the mass-dependence of small planets (both super-Earths and sub-Neptunes together) on host star mass. 
Analysis of 1841 confirmed \Kepler{} planets (using \textit{Gaia}-derived stellar mass and planet radii) finds a positive correlation for both super-Earth and sub-Neptune population's radius with stellar mass, at a slope of $\alpha = 0.23-0.35$, indicating that there is a linear relationship between planetary core mass and host star mass.
\citet{berger2020b} find agreement and measure the slope of the radius valley in radius-stellar mass space to be $\alpha=0.26^{+0.21}_{-0.16}$.
\citet{petigura2022} looked at the radius valley as a function of stellar mass in the CKS survey. They do not find a constant slope for the radius valley in radius-stellar mass space, instead finding that the super-Earth population stays constant over stellar mass and that the sub-Neptune population grows to larger radii with stellar mass. The positive correlation between planet size and stellar mass for the sub-Neptunes implies that larger stars produce larger planet cores, agreeing with \citet{wu2019}.
\citet{ho2023} use \Kepler{} short cadence observations to find a slope of $\alpha=0.231^{+0.053}_{-0.06}$.
It should be noted that these analyses are in the \textit{observed} density of planets, and not in the occurrence of planets. Observed versus intrinsic populations could explain why we see a shift in the lower edge of the radius valley where they see that the super-Earth population is constant.

\begin{figure}
    \centering
    \includegraphics[width=\columnwidth]{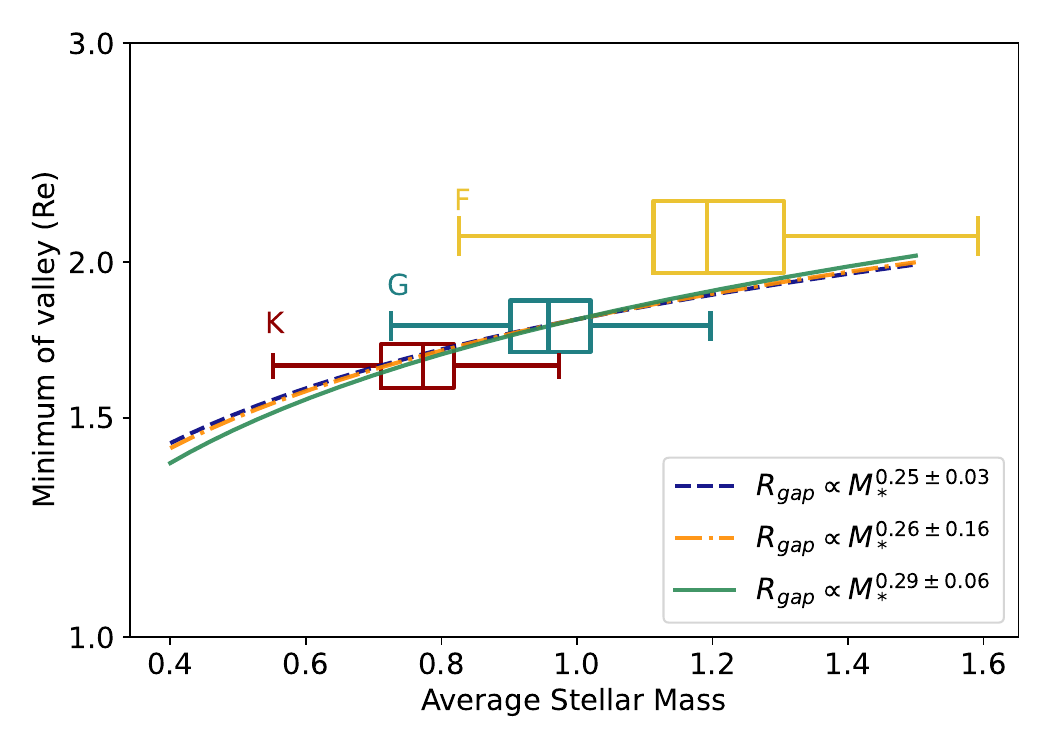}
    \caption{Stellar mass dependence of the radius valley. We do not compute occurrence as a function of stellar mass but plot box plots of the mass distribution for each stellar type. The center line of the box plot represents the median stellar mass, with first and second quartiles being marked by the edges of the box. The whiskers extend to the minimum and maximum stellar mass values. The height of the box is the uncertainty in the measured minimum of the radius valley integrated from $1-30$ days. Observed fits to the slope of the radius valley in the planet radius-stellar mass plane are plotted; our measurements are in agreement.}
    \label{fig:box_obs}
\end{figure}

We do not directly measure the slope of the valley as a function of stellar mass-planet radius, but can make an approximation by looking at the average stellar mass within each stellar type bin as a function of the measured minimum of the radius valley integrated from $1-30$ days, plotted in Figure~\ref{fig:box_obs}. Our measurements are in good agreement with the previous studies. We further compare this relationship with theoretical models in Section~\ref{sec:rv_disc}.

\subsection{The Occurrence Cliff}\label{sec:cliff}
We pay special attention to the occurrence cliff-- the sharp decline in occurrence between $3-4$ \Re{}. The feature is one of the most readily apparent features in the planet distribution. However, it is the focal point of very few theoretical studies.
Evolutionary models focusing on the radius valley often yield an occurrence cliff qualitatively similar to that observed by \Kepler{}, yet the physical processes and tuning parameters required to yield a steep cliff are not described in detail. 

Before correcting for completeness, the cliff appears to run along a constant radius contour. However, the slope of the occurrence cliff is not the same at every orbital period: the cliff extends to larger radii for longer orbital periods; i.e. the slope of the occurrence cliff becomes flatter with increasing orbital period. This phenomenon has been noticed as early as \citet{dong2013}, who noticed the ``fast rise of the Neptunes'' in early \Kepler{} data. 

\begin{figure*}
    \centering
    \includegraphics[width=0.48\textwidth]{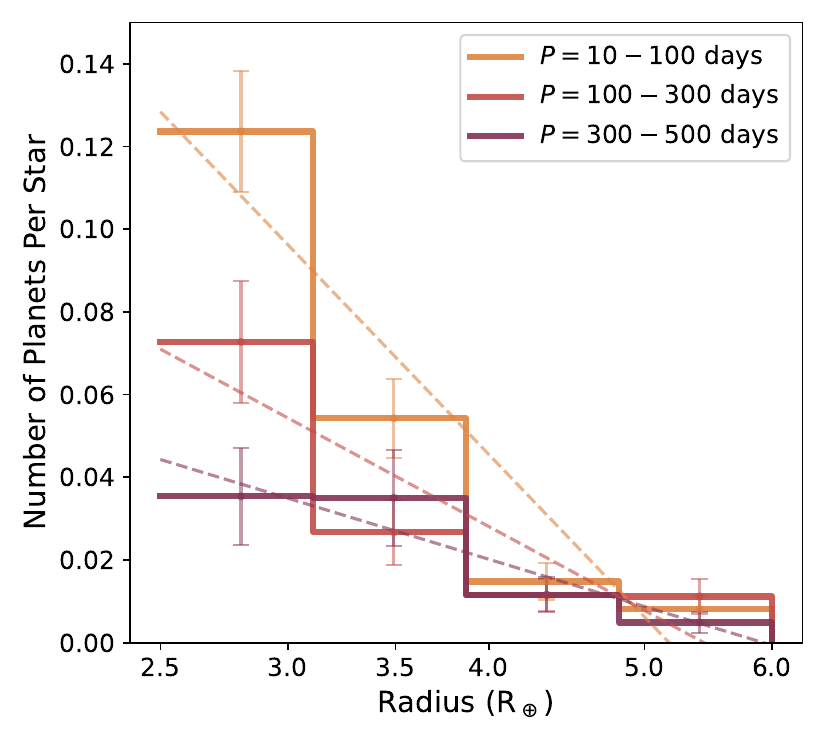}
    \includegraphics[width=0.48\textwidth]{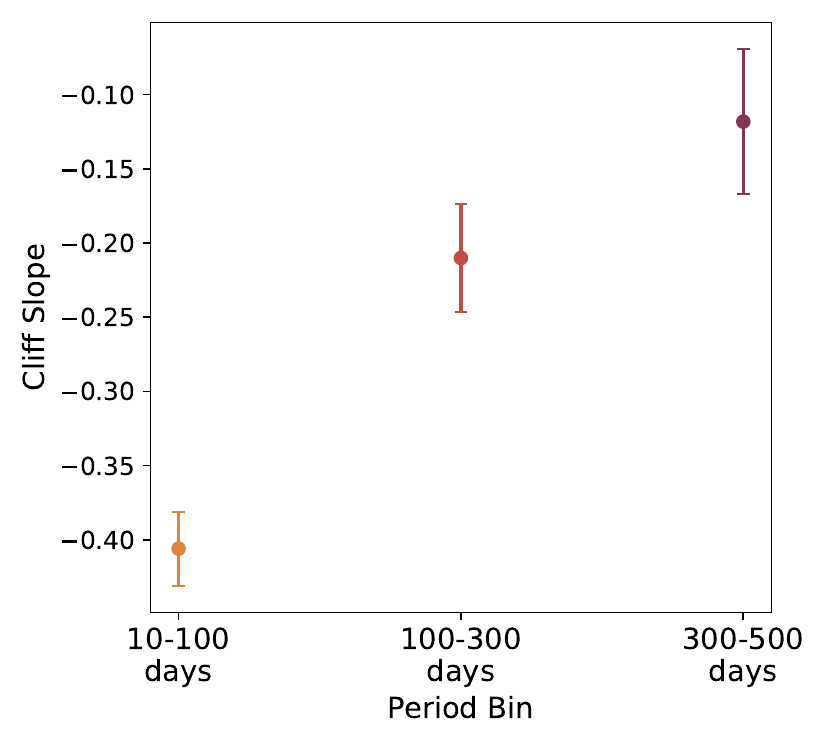}
    \caption{Left: Radius distribution for planets between 2.5-6 \Re{} separated into three different period bins, low (10- 100 days, orange); mid (100-300 days, pink); high (300-500 days, purple). The occurrence of short-period planets is steeper than the occurrence of planets at long orbital period. Right: Distribution of occurrence cliff slopes for three different period bins, measured for 1000 Monte Carlo runs. The shortest period bin, 10-100 days (orange), has the steepest slope of $m_C==-0.406^{+0.012}_{0.012}$, the middle period bin, 100-300 days (pink) has a slope of $m_C=-0.210^{+0.018}_{-0.017}$, and the longest period bin, 300-500 days (purple), has the flattest slope at $m_C=-0.121^{+0.026}_{-0.21}2$.}
    \label{fig:cliff}
\end{figure*}

We inspect this effect in Figure~\ref{fig:cliff} by looking at the radius distribution. Instead of looking only at occurrence below 100 days, we break the radius distribution into three period bins: low ($10-100$ days), mid ($100-300$ days), and high ($300-500$ days).\footnote{The longest period bin goes out to 500 days (instead of 400 as all other occurrence measurements) to cover the largest area possible to compare with the other period bins. The impacts of the completeness degradation discussed in Section~\ref{sec:completness} affect smaller radii more strongly than this regime.}
We measure the slope of the cliff between $2.5-6$ \Re{} for each bin and find that the occurrence cliff becomes gentler (with a slope closer to zero) for the longer period bins. 

We measure the slope of each period bin with the equation\footnote{Note that we are finding the slope of the integrated occurrence here, not the occurrence rate as in Eq.~\ref{eq:period}.}:
\begin{equation}\label{eq:cliff}
    \textrm{NPPS} = F = m_C \log{R_{\oplus}} + b
\end{equation}\\
where $m_c$ is the slope of occurrence cliff and $b$ is an intercept point. Similar to the distributions fit to the radius gap, we fit each iteration of our bootstrap simulation to measure the slope of each period bin. For our FGK sample, we measure mean slopes of $m_C=-0.41^{+0.01}_{-0.01}$ for $P=10-100$ days, $m_C=-0.21^{+0.02}_{-0.02}$ for $P=100-300$ days, and $m_C=-0.12^{+0.03}_{-0.21}$ for $P=300-500$ days. The right panel of Figure~\ref{fig:cliff} shows the measured slopes for each period bin. The slope of the cliff gets progressively flatter for longer orbital periods. There are no similar measurements in the literature so we do not compare here.

\subsubsection{Occurrence cliff as a function of stellar type}

\begin{figure}
    \centering
    \includegraphics[width=\columnwidth]{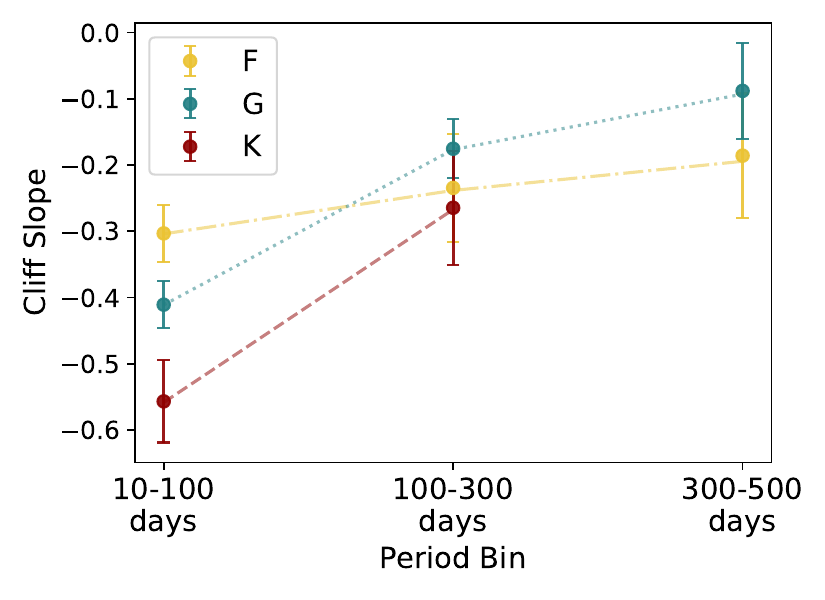}
    \caption{Slope of the occurrence cliff for each stellar type. For each stellar type, the slope of the occurrence cliff becomes flatter for longer orbital period. The K-(red) and G-star (teal) occurrence have similar changes in cliff slope; the F-star (yellow) occurrence is more similar from low to high orbital period. We exclude the last bin for K-stars because it contains only 2 planet detections.}
    \label{fig:stellar_cliff}
\end{figure}

Similar to the radius valley, we can also look at how the occurrence cliff changes over stellar type. As in Section~\ref{sec:cliff}, we break the radius distribution into three period bins: low ($10-100$ days), mid ($100-300$ days), and high ($300-500$ days). We measure the slope of each period bin for each stellar type with the same process and linear function as for the FGK sample. Table~\ref{tab:cliff} reports the measured mean slopes and $68.3\%$ confidence intervals, as well as the number of planets within each bin.
Figure~\ref{fig:stellar_cliff} shows the measured cliff slopes for each stellar type. Each stellar type follows the same trend of the shortest period bin having the steepest slope, with the slope becoming flatter with each progressive period bin. The slopes become more similar with longer orbital period for each stellar type. However, the rate of change of the slope appears to be more gradual for F stars compared to G and K stars.
We exclude the cliff slope measurements for K occurrence for the longest period bin because there are only 2 planets that reside within it. 

\begin{deluxetable*}{c|ccc}

\tabletypesize{\footnotesize}
\tablecolumns{4}
\tablewidth{0pt}
\tablecaption{ Occurrence Cliff Slope Fits \label{tab:cliff}}
\tablehead{\colhead{} & \colhead{10-100 days} & \colhead{100-300 days} & \colhead{300-500 days}}

\startdata \\[-3mm]
    FGK & $-0.406^{+0.01}_{-0.01}$ (447) & $-0.210^{+0.02}_{-0.02}$ (75) & $-0.121^{+0.03}_{-0.02}$ (16) \\[2mm]
    F & $-0.304^{+0.02}_{-0.02}$ (122) & $-0.238^{+0.04}_{-0.04}$ (24) & $-0.194^{+0.05}_{-0.04}$ (5) \\[2mm]
    G & $-0.411^{+0.02}_{-0.02}$ (218) & $-0.177^{+0.02}_{-0.02}$ (32) & $-0.093^{+0.04}_{-0.03}$ (9) \\[2mm]
    K & $-0.559^{+0.03}_{-0.03}$ (107) & $-0.268^{+0.05}_{-0.04}$ (19) & \textit{na} (2) \\[2mm]
\enddata
\tablecomments{Measured cliff slopes for each period bin and stellar type. The parenthetical numbers beside each slope measurement is the number of planets inside that bin.}

\end{deluxetable*}

\section{Discussion}\label{sec:discussion}
Our KDE methodology gives us the ability to measure structure in multiple dimensions that we cannot with parametric forms. It is flexible for comparisons with previous studies and incorporates the best-of-our knowledge input parameters in a uniform manner. There are limitations to this methodology, though. We cannot extrapolate occurrence rates to areas of parameter space with low or no planet detections, such as the habitable zone.

It is rarely possible to reproduce independent experiments exactly, but we have tested a variety of benchmark measurements across a broad range of parameter space and find excellent agreement with previous studies. Our measurements for total occurrence, hot Jupiter occurrence, and the radius distribution are all within 1-$\sigma$ of previous studies.

We find the most tension with regards to small radius and/or long period planets, such as the turn over of the super-Earth population or the slope of their period distribution. This is to be expected: there are few planet detections in these regions, and both completeness and reliability are at their lowest. These planets require very careful treatment to resolve disagreement.

\subsection{What causes the shift in the radius valley?}\label{sec:rv_disc}
When looking at the small planet radius distribution separated by stellar type, there is a distinct shift in the minimum of the radius valley to smaller radii for cooler stars. This shift could be caused by formation or evolutionary processes.

The first possible explanation is that hotter stars can strip the volatile envelopes of larger cores. According to atmospheric mass loss models, the bottom edge of the valley corresponds to bare rocky cores, and the top edge corresponds to cores with $\sim$4\% H-He atmospheres \citep{owen2017}. If the XUV flux from the star is strong enough, larger cores are completely stripped, pushing the lower edge of the valley to higher radii. There is a slight dependence of integrated lifetime XUV with stellar mass \citep{mcdonald2019}, so hotter stars could indeed photoevaporate planets more easily than cooler stars. 

Core-powered mass-loss also causes planets to lose their primordial envelopes, primarily through luminosity from the cooling rocky cores \citet{ginzburg2018}. A mechanism solely caused by the planet, independent of the stellar host environment, could not explain the shift in the radius valley that we see. However, core-powered mass-loss has a dependence on the bolometric flux from the host star (via equilibrium temperature, which sets cooling and mass loss rates), and thus cannot be ruled out.

In order to disentangle these mechanisms, we need to look at the distribution of planets as a function of planet radius, insolation flux, and stellar mass simultaneously. Some recent work \citep[e.g.][]{rogers2021, berger2023} has begun to study this, but we will defer studying it in occurrence space to a future paper.

\begin{figure}[t]
    \centering
    \includegraphics[width=\columnwidth]{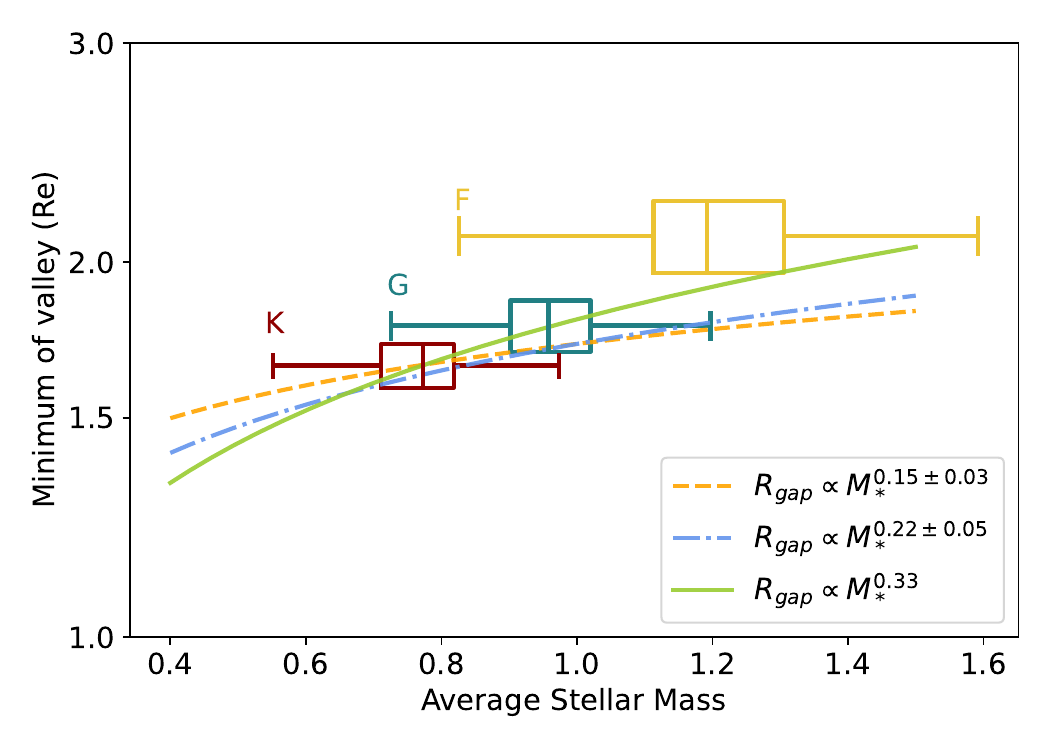}
    \caption{Dependence of the radius valley as a function of stellar mass. We do not measure occurrence as a function of stellar mass; here we investigate this dependence by plotting a box plot to represent the distribution of stellar masses for each stellar type. The center line for each box plot represents the median stellar mass value. The left and right edges of the box show one quartile above and below the median, while the whiskers show the minimum and maximum mass values. The width of the box is the uncertainty in the radius valley minimum measurement integrated from $1-30$ days. We show power-law fits to the radius valley from \citet{lee2022}: the bottom-heavy core mass model ($R_{gap} \propto M_*^{0.22 \pm 0.05}$) in blue and the top-heavy core mass model ($R_{gap} \propto M_*^{0.15 \pm 0.03}$) in orange, as well as the power-law derived from core-powered mass-loss models ($R_{gap} \propto M_*^{0.33}$ \citep{gupta2020}). Our results are most consistent with the steeper slope of the core-powered mass-loss model.}
	\label{fig:mass_min}
\end{figure}

There could also be a formation cause for the shift in the valley. \citet{lee2022} theorized that with more realistic disk physics, the radius valley can be created without any atmospheric mass loss processes by gas accretion alone. There is a  maximal isothermal envelope mass each planet is able to accrete according to how massive the planet core is and its disk location. Small cores, 2-3M$_\oplus$, are unable to accrete any atmosphere and remain bare, while massive cores $>$ 10 M$_\oplus$ undergo runaway accretion before being halted by disk dispersal. The isothermal-to-cooling transition carves out a gap at $\sim$1.2 \Re{} in the radius distribution, replicating the observed radius valley. 

This phenomenon should replicate the trend we see with a shifted radius valley with both stellar temperature and mass. We take the average stellar mass within each stellar type for our sample and plot it against our measured minimum of the radius valley from $1-30$ days, seen in Figure~\ref{fig:mass_min}. We can compare this to the power laws \citet{lee2022} established for bottom-heavy (blue) and top-heavy (orange) core mass distributions, as seen in Figure 12 from \citet{lee2022}. This theoretical fit for the gap agrees with those found in \cite{petigura2022} and \citet{berger2020b}. As with the comparison with the observational fits in Figure~\ref{fig:box_obs}, our measurements show good agreement with the models. However, a steeper slope like the core-powered mass-loss model \citep[][green]{gupta2020} is preferred.

\subsection{Change in the slope of the Occurrence Cliff}\label{sec:cliff_disc}
The same thing that shifts the location of the radius valley can shape the occurrence cliff, as it is often an outcome of models that replicate the radius valley. The change in cliff slope across this wide span in period could be tied to different physical processes. If we assume the main drivers of the radius valley-- photoevaporation and core-powered mass loss-- primarily work within 100 days, the difference in slopes for the low- and mid-period bins (10-100 days vs. 100-300 days) can be attributed to the drop-off in efficacy in those physical processes. As a planet becomes less irradiated and cooler at longer orbital period, it can potentially retain more of its envelope and sit at a larger radius. As a population, a wider range of radii causes the cliff slope to be gentler.

If the drivers of the radius valley mainly act within 100 days, they cannot explain the continued trend in gentler cliff slopes we see beyond 300 days. Instead, this more closely matches the prediction from \citet{kite2019}, where they theorized that the cause of the occurrence cliff is the ``fugacity crisis,'' where the radii of sub-Neptune-sized planets are limited by the sequestration of hydrogen in the planet's magma ocean. 
This mechanism is less efficient, on average, for cooler planet temperatures. The implication is that planets at longer orbital periods tend to maintain larger radii, thereby reducing the slope of the occurrence cliff.

The changing slope of the cliff could also trace a changing core-mass distribution for small planets. As orbital period (and therefore semi-major axis) increases, the availability of materials in the disk increases, which could lead to increased planet core sizes.

\section{Summary \& Conclusion}\label{sec:conclusions}
In this paper, we:
\begin{itemize}
    \item present a KDE methodology for measuring \Kepler{} occurrence rates. We incorporate various measurement uncertainties, \textit{Gaia} DR2 stellar radii, and reliability measurements into the calculations.
    \item validate the method with a toy model and quantify the level of structure we can measure, KDE edge effects, and a low completeness boundary where we are unable to measure occurrence below.
    \item measure the period and radius distribution of the \Kepler{} FGK sample and compare to previous studies. We find a total FGK occurrence of $1.52 \pm 0.08$ NPPS between $1-400$ days and $0.5-16$ \Re{}.
    \item fit the period distribution to a parametric distribution from \citet{howard2012} and find that our fit agrees with previous measurements.
    \item measure the shape of the small planet radius distribution. We find the minimum of the radius valley to be $1.78^{+0.14}_{-0.16}$ \Re{}.
    \item fit the period distribution to a parametric distribution from \citet{howard2012} and find that our fit agrees with previous measurements. 
    \item measure occurrence dependent on stellar type and confirm that occurrence increases towards cooler stars. We find total occurrences (P$=1-400$ days; R$=0.5-16$ \Re{}) of $1.05\pm0.09$ NPPS for F stars, $1.26\pm0.09$ NPPS for G stars, and $1.61\pm0.13$ NPPS for K stars.
    \item find that the minimum of the radius valley moves to smaller radii for cooler stars, from $1.92^{+0.26}_{-0.18}$ \Re{} for F stars to to $1.80^{+0.16}_{-0.05}$ \Re{} for G stars $1.62^{+0.02}_{-0.04}$ \Re{} for K stars.
    \item measure the slope of the occurrence cliff and find it to be less steep at longer orbital period. There is also a dependence on stellar type: the total change between short and long orbital period bins for cliff slope becomes larger for cooler stars.
\end{itemize}

Our methodology enables us to map the occurrence structure to physical models of planet formation and evolution theory because we can measure any type of underlying structure in planet occurrence. Throughout Section~\ref{sec:occ}, we pointed to measurements that could be directly tied to theories. The next step is to use this framework to compare these measurements with populations derived from models of planet formation and evolution.

The features we look at in this paper, namely the radius valley and occurrence cliff, will be more informative in insolation rather than orbital period, to mitigates the effects of varying stellar biases. We reserve this work for a future paper.

Utilizing multidimensional occurrence measurements will allow us to constrain input parameters for planet formation and evolution. In the future, we will use population synthesis models in a Bayesian framework to evolve planets through physical models and measure their occurrence to see if they match our actual measurements. This will enable us not only to constrain free parameters in the theoretical models but also to compare theoretical models to each other.

\section*{Acknowledgments}
    The authors would like to thank the anonymous referee whose comments helped improve the paper.
    AD would like to thank Michelle Kunimoto, Danley Hsu, Kendall Sullivan, Ruth Murray-Clay, James Rogers, Joey Murphy, Jonathan Fortney, and Adam Kraus for helpful conversations in the completion of this manuscript. 
    AD gratefully acknowledges support from the Heising-Simons Foundation through grant 2021-3197.
    We acknowledge use of the lux supercomputer at UC Santa Cruz, funded by NSF MRI grant AST 1828315.
    This work used the Extreme Science and Engineering Discovery Environment (XSEDE) \textit{Expanse} at the San Diego Supercomputer Center through allocation TG-PHY210033 \citep{xsede}.

\software{Additional packages used include: \textit{Astropy} \citep{astropy}, \textit{numpy} \citep{numpy}, \textit{emcee} \citep{emcee}, \textit{pandas} \citep{pandas}, \textit{scipy} \citep{scipy}, \textit{matplotlib} \citep{matplotlib}}

\appendix
\section{Vetting Completeness}\label{ap:vet}
Following \citet{bryson2020a}, we grid period-expected MES space and treat the fraction of recovered PCs in each cell as a binomial rate. We fit the rate $\rho$ with a product of a non-rotated simplified logisitic function in period $p$ times a rotated logistic in $p$ and expected MES $m$. This gives us 6 parameters to fit with our Bayseian inference: $\theta = [x_0, y_0, k_x, k_y, \phi, A]$, where:
\begin{align*}
    x &= \frac{(p - p_{min})}{p_{max} - p_{min}}\\
    y &= \frac{(m - m_{min})}{m_{max} - m_{min}} \\
    y_{rot} &= (y - 0.5) \cdot \cos{\phi} - (x - 0.5) \cdot \sin{\phi} \\
    \rho &= A \; Y(x,\, x_0,\, -k_x,\, 1) \times Y(y_{rot} + 0.5,\, y_0,\, k_y,\, 1)
\end{align*}

Our MCMC computation used 100 walkers and ran for 10000 steps. We ensured convergence by checking that the number of steps were greater than 50x the estimate of the integrated autocorrelation time. The maximum parameter was $\tau=92$.

For our FGK stellar sample, the MCMC median posteriors and 84th and 16th percentiles are:

\smallskip
$x_0 = 1.040^{+0.035}_{-0.031}$,
$y_0 = 0.153^{+0.006}_{-0.007}$,

$k_x = 5.055^{+0.501}_{-0.489}$,
$k_y = 18.877^{+1.320}_{-1.283}$,

$\phi = 4.019^{+0.911}_{-0.875}$,
$A = 0.964^{+0.005}_{-0.005}$
\smallskip

Our separate F, G, and K inferences are all within 1-$\sigma$ uncertainties  for the FGK fits so we do not report them here.

\section{Reliability}\label{ap:rel}
\subsection{False Alarm Effectiveness}
We follow the same process as vetting completeness, following \citet{bryson2020a}, to characterize the false alarm effectiveness, instead fitting the binomial rate $\rho$ with a simple rotated logistic function for $theta=[x_0, k_x, \phi, A]$:

\begin{align*}
    x &= \frac{(p - p_{min})}{p_{max} - p_{min}}\\
    y &= \frac{(m - m_{min})}{m_{max} - m_{min}} \\
    x_{rot} &= (x - 0.5) \cdot \cos{\phi} - (y - 0.5) \cdot \sin{\phi} \\
    \rho &= A \; Y(x_{rot} + 0.5,\, x_0,\, k_y,\, 1)
\end{align*}

Our MCMC computation used 100 walkers and ran for 10000 steps. We ensured convergence by checking that the number of steps were greater than 50x the estimate of the integrated autocorrelation time. The maximum parameter was $\tau=197$.
For our FGK stellar sample, the MCMC median posteriors and 84th and 16th percentiles are:

\smallskip
$x_0 = 1.079^{+0.034}_{-0.025}$, 
$k_x = 51.6292^{+4.585-}_{-15.666}$,

$A = 0.996^{+0.001}_{-0.001}$, 
$\phi = 91.151^{+0.694}_{-0.529}$
\smallskip

Our separate F, G, and K inferences are all within 1-$\sigma$ uncertainties  for the FGK fits so we do not report them here.

\subsection{Observed False Positive Rate}
We follow the same process as the false alarm effectiveness, following \citet{bryson2020a}, to characterize the observed false positive rate, instead fitting the binomial rate $\rho$ with a simple rotated logistic function for $theta=[x_0, k_x, \phi, A]$:

\begin{align*}
    x &= \frac{(p - p_{min})}{p_{max} - p_{min}}\\
    y &= \frac{(m - m_{min})}{m_{max} - m_{min}} \\
    x_{rot} &= (x - 0.5) \cdot \cos{\phi} - (y - 0.5) \cdot \sin{\phi} \\
    \rho &= A \; Y(x_{rot} + 0.5,\, x_0,\, k_y,\, 1)
\end{align*}

Our MCMC computation used 100 walkers and ran for 10000 steps. We ensured convergence by checking that the number of steps were greater than 50x the estimate of the integrated autocorrelation time. The maximum parameter was $\tau= 154$.

For our FGK stellar sample, the MCMC median posteriors and 84th and 16th percentiles are:

\smallskip
$x_0 = 0.639^{+0.019}_{-0.019}$,
$k_x = 9.320^{+0.411}_{-0.377}$,

$A = 0.991^{+0.004}_{-0.005}$,
$\phi = -157.457^{+2.039}_{-2.012}$
\smallskip

Our separate F, G, and K inferences are all within 1-$\sigma$ uncertainties for the FGK fits so we do not report them here.

\bibliography{sample631}{}

\begin{thebibliography}{}
\expandafter\ifx\csname natexlab\endcsname\relax\def\natexlab#1{#1}\fi
\providecommand{\url}[1]{\href{#1}{#1}}
\providecommand{\dodoi}[1]{doi:~\href{http://doi.org/#1}{\nolinkurl{#1}}}
\providecommand{\doeprint}[1]{\href{http://ascl.net/#1}{\nolinkurl{http://ascl.net/#1}}}
\providecommand{\doarXiv}[1]{\href{https://arxiv.org/abs/#1}{\nolinkurl{https://arxiv.org/abs/#1}}}

\bibitem[{Akaike(1974)}]{akaike1974}
Akaike, H. 1974, IEEE Transactions on Automatic Control, 19, 716,
  \dodoi{10.1109/TAC.1974.1100705}

\bibitem[{{Astropy Collaboration} {et~al.}(2013){Astropy Collaboration},
  {Robitaille}, {Tollerud}, {Greenfield}, {Droettboom}, {Bray}, {Aldcroft},
  {Davis}, {Ginsburg}, {Price-Whelan}, {Kerzendorf}, {Conley}, {Crighton},
  {Barbary}, {Muna}, {Ferguson}, {Grollier}, {Parikh}, {Nair}, {Unther},
  {Deil}, {Woillez}, {Conseil}, {Kramer}, {Turner}, {Singer}, {Fox}, {Weaver},
  {Zabalza}, {Edwards}, {Azalee Bostroem}, {Burke}, {Casey}, {Crawford},
  {Dencheva}, {Ely}, {Jenness}, {Labrie}, {Lim}, {Pierfederici}, {Pontzen},
  {Ptak}, {Refsdal}, {Servillat}, \& {Streicher}}]{astropy}
{Astropy Collaboration}, {Robitaille}, T.~P., {Tollerud}, E.~J., {et~al.} 2013,
  \aap, 558, A33, \dodoi{10.1051/0004-6361/201322068}

\bibitem[{{Beaug{\'e}} \& {Nesvorn{\'y}}(2013)}]{beauge2013}
{Beaug{\'e}}, C., \& {Nesvorn{\'y}}, D. 2013, \apj, 763, 12,
  \dodoi{10.1088/0004-637X/763/1/12}

\bibitem[{{Beleznay} \& {Kunimoto}(2022)}]{beleznay2022}
{Beleznay}, M., \& {Kunimoto}, M. 2022, \mnras, 516, 75,
  \dodoi{10.1093/mnras/stac2179}

\bibitem[{{Berger} {et~al.}(2018){Berger}, {Huber}, {Gaidos}, \& {van
  Saders}}]{berger2018}
{Berger}, T.~A., {Huber}, D., {Gaidos}, E., \& {van Saders}, J.~L. 2018, \apj,
  866, 99, \dodoi{10.3847/1538-4357/aada83}

\bibitem[{{Berger} {et~al.}(2020{\natexlab{a}}){Berger}, {Huber}, {Gaidos},
  {van Saders}, \& {Weiss}}]{berger2020b}
{Berger}, T.~A., {Huber}, D., {Gaidos}, E., {van Saders}, J.~L., \& {Weiss},
  L.~M. 2020{\natexlab{a}}, \aj, 160, 108, \dodoi{10.3847/1538-3881/aba18a}

\bibitem[{{Berger} {et~al.}(2020{\natexlab{b}}){Berger}, {Huber}, {van Saders},
  {Gaidos}, {Tayar}, \& {Kraus}}]{berger2020a}
{Berger}, T.~A., {Huber}, D., {van Saders}, J.~L., {et~al.} 2020{\natexlab{b}},
  \aj, 159, 280, \dodoi{10.3847/1538-3881/159/6/280}

\bibitem[{{Berger} {et~al.}(2023){Berger}, {Schlieder}, {Huber}, \&
  {Barclay}}]{berger2023}
{Berger}, T.~A., {Schlieder}, J.~E., {Huber}, D., \& {Barclay}, T. 2023, arXiv
  e-prints, arXiv:2302.00009, \dodoi{10.48550/arXiv.2302.00009}

\bibitem[{{Borucki} {et~al.}(2010){Borucki}, {Koch}, {Basri}, {Batalha},
  {Brown}, {Caldwell}, {Caldwell}, {Christensen-Dalsgaard}, {Cochran},
  {DeVore}, {Dunham}, {Dupree}, {Gautier}, {Geary}, {Gilliland}, {Gould},
  {Howell}, {Jenkins}, {Kondo}, {Latham}, {Marcy}, {Meibom}, {Kjeldsen},
  {Lissauer}, {Monet}, {Morrison}, {Sasselov}, {Tarter}, {Boss}, {Brownlee},
  {Owen}, {Buzasi}, {Charbonneau}, {Doyle}, {Fortney}, {Ford}, {Holman},
  {Seager}, {Steffen}, {Welsh}, {Rowe}, {Anderson}, {Buchhave}, {Ciardi},
  {Walkowicz}, {Sherry}, {Horch}, {Isaacson}, {Everett}, {Fischer}, {Torres},
  {Johnson}, {Endl}, {MacQueen}, {Bryson}, {Dotson}, {Haas}, {Kolodziejczak},
  {Van Cleve}, {Chandrasekaran}, {Twicken}, {Quintana}, {Clarke}, {Allen},
  {Li}, {Wu}, {Tenenbaum}, {Verner}, {Bruhweiler}, {Barnes}, \&
  {Prsa}}]{borucki2010}
{Borucki}, W.~J., {Koch}, D., {Basri}, G., {et~al.} 2010, Science, 327, 977,
  \dodoi{10.1126/science.1185402}

\bibitem[{{Borucki} {et~al.}(2011){Borucki}, {Koch}, {Basri}, {Batalha},
  {Brown}, {Bryson}, {Caldwell}, {Christensen-Dalsgaard}, {Cochran}, {DeVore},
  {Dunham}, {Gautier}, {Geary}, {Gilliland}, {Gould}, {Howell}, {Jenkins},
  {Latham}, {Lissauer}, {Marcy}, {Rowe}, {Sasselov}, {Boss}, {Charbonneau},
  {Ciardi}, {Doyle}, {Dupree}, {Ford}, {Fortney}, {Holman}, {Seager},
  {Steffen}, {Tarter}, {Welsh}, {Allen}, {Buchhave}, {Christiansen}, {Clarke},
  {Das}, {D{\'e}sert}, {Endl}, {Fabrycky}, {Fressin}, {Haas}, {Horch},
  {Howard}, {Isaacson}, {Kjeldsen}, {Kolodziejczak}, {Kulesa}, {Li}, {Lucas},
  {Machalek}, {McCarthy}, {MacQueen}, {Meibom}, {Miquel}, {Prsa}, {Quinn},
  {Quintana}, {Ragozzine}, {Sherry}, {Shporer}, {Tenenbaum}, {Torres},
  {Twicken}, {Van Cleve}, {Walkowicz}, {Witteborn}, \& {Still}}]{borucki2011}
{Borucki}, W.~J., {Koch}, D.~G., {Basri}, G., {et~al.} 2011, \apj, 736, 19,
  \dodoi{10.1088/0004-637X/736/1/19}

\bibitem[{{Bressan} {et~al.}(2012){Bressan}, {Marigo}, {Girardi}, {Salasnich},
  {Dal Cero}, {Rubele}, \& {Nanni}}]{bressan2012}
{Bressan}, A., {Marigo}, P., {Girardi}, L., {et~al.} 2012, \mnras, 427, 127,
  \dodoi{10.1111/j.1365-2966.2012.21948.x}

\bibitem[{{Bryan} {et~al.}(2016){Bryan}, {Knutson}, {Howard}, {Ngo}, {Batygin},
  {Crepp}, {Fulton}, {Hinkley}, {Isaacson}, {Johnson}, {Marcy}, \&
  {Wright}}]{bryan2016}
{Bryan}, M.~L., {Knutson}, H.~A., {Howard}, A.~W., {et~al.} 2016, \apj, 821,
  89, \dodoi{10.3847/0004-637X/821/2/89}

\bibitem[{{Bryson} {et~al.}(2020{\natexlab{a}}){Bryson}, {Coughlin}, {Batalha},
  {Berger}, {Huber}, {Burke}, {Dotson}, \& {Mullally}}]{bryson2020a}
{Bryson}, S., {Coughlin}, J., {Batalha}, N.~M., {et~al.} 2020{\natexlab{a}},
  \aj, 159, 279, \dodoi{10.3847/1538-3881/ab8a30}

\bibitem[{{Bryson} {et~al.}(2020{\natexlab{b}}){Bryson}, {Coughlin},
  {Kunimoto}, \& {Mullally}}]{bryson2020b}
{Bryson}, S., {Coughlin}, J.~L., {Kunimoto}, M., \& {Mullally}, S.~E.
  2020{\natexlab{b}}, \aj, 160, 200, \dodoi{10.3847/1538-3881/abb316}

\bibitem[{{Burke} \& {Catanzarite}(2017)}]{burke2017}
{Burke}, C.~J., \& {Catanzarite}, J. 2017, {Planet Detection Metrics:
  Per-Target Detection Contours for Data Release 25}, Kepler Science Document
  KSCI-19111-002

\bibitem[{Burke {et~al.}(2015)Burke, Christiansen, Mullally, Seader, Huber,
  Rowe, Coughlin, Thompson, Catanzarite, Clarke, Morton, Caldwell, Bryson,
  Haas, Batalha, Jenkins, Tenenbaum, Twicken, Li, Quintana, Barclay, Henze,
  Borucki, Howell, \& Still}]{burke2015}
Burke, C.~J., Christiansen, J.~L., Mullally, F., {et~al.} 2015, The
  Astrophysical Journal, 809, 8, \dodoi{10.1088/0004-637x/809/1/8}

\bibitem[{{Catanzarite} \& {Shao}(2011)}]{catanzarite2011}
{Catanzarite}, J., \& {Shao}, M. 2011, \apj, 738, 151,
  \dodoi{10.1088/0004-637X/738/2/151}

\bibitem[{{Christiansen}(2017)}]{christiansen2017}
{Christiansen}, J.~L. 2017, {Planet Detection Metrics: Pixel-Level Transit
  Injection Tests of Pipeline Detection Efficiency for Data Release 25}, Kepler
  Science Document KSCI-19110-001, id. 18. Edited by Michael R. Haas and
  Natalie M. Batalha

\bibitem[{{Christiansen} {et~al.}(2013){Christiansen}, {Clarke}, {Burke},
  {Jenkins}, {Barclay}, {Ford}, {Haas}, {Sabale}, {Seader}, {Claiborne Smith},
  {Tenenbaum}, {Twicken}, {Kamal Uddin}, \& {Thompson}}]{christiansen2013}
{Christiansen}, J.~L., {Clarke}, B.~D., {Burke}, C.~J., {et~al.} 2013, \apjs,
  207, 35, \dodoi{10.1088/0067-0049/207/2/35}

\bibitem[{{Christiansen} {et~al.}(2015){Christiansen}, {Clarke}, {Burke},
  {Seader}, {Jenkins}, {Twicken}, {Catanzarite}, {Smith}, {Batalha}, {Haas},
  {Thompson}, {Campbell}, {Sabale}, \& {Kamal Uddin}}]{christiansen2015}
---. 2015, \apj, 810, 95, \dodoi{10.1088/0004-637X/810/2/95}

\bibitem[{{Christiansen} {et~al.}(2016){Christiansen}, {Clarke}, {Burke},
  {Jenkins}, {Bryson}, {Coughlin}, {Mullally}, {Thompson}, {Twicken},
  {Batalha}, {Haas}, {Catanzarite}, {Campbell}, {Kamal Uddin}, {Zamudio},
  {Smith}, \& {Henze}}]{christiansen2016}
---. 2016, \apj, 828, 99, \dodoi{10.3847/0004-637X/828/2/99}

\bibitem[{{Coughlin} {et~al.}(2017){Coughlin}, {Thompson}, \& {Kepler
  Team}}]{coughlin2017}
{Coughlin}, J., {Thompson}, S.~E., \& {Kepler Team}. 2017, in American
  Astronomical Society Meeting Abstracts, Vol. 230, American Astronomical
  Society Meeting Abstracts \#230, 102.04

\bibitem[{{Dawson} \& {Murray-Clay}(2013)}]{dawson2013}
{Dawson}, R.~I., \& {Murray-Clay}, R.~A. 2013, \apjl, 767, L24,
  \dodoi{10.1088/2041-8205/767/2/L24}

\bibitem[{{Dong} \& {Zhu}(2013)}]{dong2013}
{Dong}, S., \& {Zhu}, Z. 2013, \apj, 778, 53,
  \dodoi{10.1088/0004-637X/778/1/53}

\bibitem[{{Dressing} \& {Charbonneau}(2013)}]{dressing2013}
{Dressing}, C.~D., \& {Charbonneau}, D. 2013, \apj, 767, 95,
  \dodoi{10.1088/0004-637X/767/1/95}

\bibitem[{{Dressing} \& {Charbonneau}(2015)}]{dressing2015}
---. 2015, \apj, 807, 45, \dodoi{10.1088/0004-637X/807/1/45}

\bibitem[{{Eker} {et~al.}(2018){Eker}, {Bak{\i}{\c{s}}}, {Bilir}, {Soydugan},
  {Steer}, {Soydugan}, {Bak{\i}{\c{s}}}, {Ali{\c{c}}avu{\c{s}}}, {Aslan}, \&
  {Alpsoy}}]{eker2018}
{Eker}, Z., {Bak{\i}{\c{s}}}, V., {Bilir}, S., {et~al.} 2018, \mnras, 479,
  5491, \dodoi{10.1093/mnras/sty1834}

\bibitem[{{Fernandes} {et~al.}(2019){Fernandes}, {Mulders}, {Pascucci},
  {Mordasini}, \& {Emsenhuber}}]{fernades2019}
{Fernandes}, R.~B., {Mulders}, G.~D., {Pascucci}, I., {Mordasini}, C., \&
  {Emsenhuber}, A. 2019, \apj, 874, 81, \dodoi{10.3847/1538-4357/ab0300}

\bibitem[{Foreman-Mackey {et~al.}(2013)Foreman-Mackey, Hogg, Lang, \&
  Goodman}]{emcee}
Foreman-Mackey, D., Hogg, D.~W., Lang, D., \& Goodman, J. 2013, Publications of
  the Astronomical Society of the Pacific, 125, 306, \dodoi{10.1086/670067}

\bibitem[{{Foreman-Mackey} {et~al.}(2014){Foreman-Mackey}, {Hogg}, \&
  {Morton}}]{dfm2014}
{Foreman-Mackey}, D., {Hogg}, D.~W., \& {Morton}, T.~D. 2014, \apj, 795, 64,
  \dodoi{10.1088/0004-637X/795/1/64}

\bibitem[{{Fressin} {et~al.}(2013){Fressin}, {Torres}, {Charbonneau}, {Bryson},
  {Christiansen}, {Dressing}, {Jenkins}, {Walkowicz}, \&
  {Batalha}}]{fressin2013}
{Fressin}, F., {Torres}, G., {Charbonneau}, D., {et~al.} 2013, \apj, 766, 81,
  \dodoi{10.1088/0004-637X/766/2/81}

\bibitem[{{Fulton} {et~al.}(2017){Fulton}, {Petigura}, {Howard}, {Isaacson},
  {Marcy}, {Cargile}, {Hebb}, {Weiss}, {Johnson}, {Morton}, {Sinukoff},
  {Crossfield}, \& {Hirsch}}]{fulton2017}
{Fulton}, B.~J., {Petigura}, E.~A., {Howard}, A.~W., {et~al.} 2017, \aj, 154,
  109, \dodoi{10.3847/1538-3881/aa80eb}

\bibitem[{{Gaidos} {et~al.}(2016){Gaidos}, {Mann}, {Kraus}, \&
  {Ireland}}]{gaidos2016}
{Gaidos}, E., {Mann}, A.~W., {Kraus}, A.~L., \& {Ireland}, M. 2016, \mnras,
  457, 2877, \dodoi{10.1093/mnras/stw097}

\bibitem[{{Ginzburg} {et~al.}(2016){Ginzburg}, {Schlichting}, \&
  {Sari}}]{ginzburg2016}
{Ginzburg}, S., {Schlichting}, H.~E., \& {Sari}, R. 2016, \apj, 825, 29,
  \dodoi{10.3847/0004-637X/825/1/29}

\bibitem[{{Ginzburg} {et~al.}(2018){Ginzburg}, {Schlichting}, \&
  {Sari}}]{ginzburg2018}
---. 2018, \mnras, 476, 759, \dodoi{10.1093/mnras/sty290}

\bibitem[{{Guo} {et~al.}(2017){Guo}, {Johnson}, {Mann}, {Kraus}, {Curtis}, \&
  {Latham}}]{guo2017}
{Guo}, X., {Johnson}, J.~A., {Mann}, A.~W., {et~al.} 2017, \apj, 838, 25,
  \dodoi{10.3847/1538-4357/aa6004}

\bibitem[{{Gupta} \& {Schlichting}(2020)}]{gupta2020}
{Gupta}, A., \& {Schlichting}, H.~E. 2020, \mnras, 493, 792,
  \dodoi{10.1093/mnras/staa315}

\bibitem[{{Hardegree-Ullman} {et~al.}(2019){Hardegree-Ullman}, {Cushing},
  {Muirhead}, \& {Christiansen}}]{khu2019}
{Hardegree-Ullman}, K.~K., {Cushing}, M.~C., {Muirhead}, P.~S., \&
  {Christiansen}, J.~L. 2019, \aj, 158, 75, \dodoi{10.3847/1538-3881/ab21d2}

\bibitem[{{Ho} \& {Van Eylen}(2023)}]{ho2023}
{Ho}, C. S.~K., \& {Van Eylen}, V. 2023, arXiv e-prints, arXiv:2301.04062.
\newblock \doarXiv{2301.04062}

\bibitem[{{Howard} {et~al.}(2012){Howard}, {Marcy}, {Bryson}, {Jenkins},
  {Rowe}, {Batalha}, {Borucki}, {Koch}, {Dunham}, {Gautier}, {Van Cleve},
  {Cochran}, {Latham}, {Lissauer}, {Torres}, {Brown}, {Gilliland}, {Buchhave},
  {Caldwell}, {Christensen-Dalsgaard}, {Ciardi}, {Fressin}, {Haas}, {Howell},
  {Kjeldsen}, {Seager}, {Rogers}, {Sasselov}, {Steffen}, {Basri},
  {Charbonneau}, {Christiansen}, {Clarke}, {Dupree}, {Fabrycky}, {Fischer},
  {Ford}, {Fortney}, {Tarter}, {Girouard}, {Holman}, {Johnson}, {Klaus},
  {Machalek}, {Moorhead}, {Morehead}, {Ragozzine}, {Tenenbaum}, {Twicken},
  {Quinn}, {Isaacson}, {Shporer}, {Lucas}, {Walkowicz}, {Welsh}, {Boss},
  {Devore}, {Gould}, {Smith}, {Morris}, {Prsa}, {Morton}, {Still}, {Thompson},
  {Mullally}, {Endl}, \& {MacQueen}}]{howard2012}
{Howard}, A.~W., {Marcy}, G.~W., {Bryson}, S.~T., {et~al.} 2012, \apjs, 201,
  15, \dodoi{10.1088/0067-0049/201/2/15}

\bibitem[{{Hsu} {et~al.}(2019){Hsu}, {Ford}, {Ragozzine}, \& {Ashby}}]{hsu2019}
{Hsu}, D.~C., {Ford}, E.~B., {Ragozzine}, D., \& {Ashby}, K. 2019, \aj, 158,
  109, \dodoi{10.3847/1538-3881/ab31ab}

\bibitem[{Hunter(2007)}]{matplotlib}
Hunter, J.~D. 2007, Computing in Science \& Engineering, 9, 90,
  \dodoi{10.1109/MCSE.2007.55}

\bibitem[{{Jenkins}(2002)}]{jenkins2002}
{Jenkins}, J.~M. 2002, \apj, 575, 493, \dodoi{10.1086/341136}

\bibitem[{{Jin}(2021)}]{jin2021}
{Jin}, S. 2021, \mnras, 502, 5302, \dodoi{10.1093/mnras/stab436}

\bibitem[{{Kane} {et~al.}(2014){Kane}, {Kopparapu}, \&
  {Domagal-Goldman}}]{kane2014}
{Kane}, S.~R., {Kopparapu}, R.~K., \& {Domagal-Goldman}, S.~D. 2014, \apjl,
  794, L5, \dodoi{10.1088/2041-8205/794/1/L5}

\bibitem[{{Kite} {et~al.}(2019){Kite}, {Fegley}, {Schaefer}, \&
  {Ford}}]{kite2019}
{Kite}, E.~S., {Fegley}, Bruce, J., {Schaefer}, L., \& {Ford}, E.~B. 2019,
  \apjl, 887, L33, \dodoi{10.3847/2041-8213/ab59d9}

\bibitem[{{Kopparapu}(2013)}]{kopparapu2013}
{Kopparapu}, R.~K. 2013, \apjl, 767, L8, \dodoi{10.1088/2041-8205/767/1/L8}

\bibitem[{{Kunimoto} \& {Matthews}(2020)}]{kunimoto2020b}
{Kunimoto}, M., \& {Matthews}, J.~M. 2020, \aj, 159, 248,
  \dodoi{10.3847/1538-3881/ab88b0}

\bibitem[{{Kunimoto} {et~al.}(2020){Kunimoto}, {Matthews}, \&
  {Ngo}}]{kunimoto2020a}
{Kunimoto}, M., {Matthews}, J.~M., \& {Ngo}, H. 2020, \aj, 159, 124,
  \dodoi{10.3847/1538-3881/ab6cf8}

\bibitem[{{Lee} \& {Chiang}(2016)}]{Lee2016}
{Lee}, E.~J., \& {Chiang}, E. 2016, \apj, 817, 90,
  \dodoi{10.3847/0004-637X/817/2/90}

\bibitem[{{Lee} {et~al.}(2022){Lee}, {Karalis}, \& {Thorngren}}]{lee2022}
{Lee}, E.~J., {Karalis}, A., \& {Thorngren}, D.~P. 2022, arXiv e-prints,
  arXiv:2201.09898.
\newblock \doarXiv{2201.09898}

\bibitem[{{Lindegren} {et~al.}(2018){Lindegren}, {Hern{\'a}ndez}, {Bombrun},
  {Klioner}, {Bastian}, {Ramos-Lerate}, {de Torres}, {Steidelm{\"u}ller},
  {Stephenson}, {Hobbs}, {Lammers}, {Biermann}, {Geyer}, {Hilger}, {Michalik},
  {Stampa}, {McMillan}, {Casta{\~n}eda}, {Clotet}, {Comoretto}, {Davidson},
  {Fabricius}, {Gracia}, {Hambly}, {Hutton}, {Mora}, {Portell}, {van Leeuwen},
  {Abbas}, {Abreu}, {Altmann}, {Andrei}, {Anglada}, {Balaguer-N{\'u}{\~n}ez},
  {Barache}, {Becciani}, {Bertone}, {Bianchi}, {Bouquillon}, {Bourda},
  {Br{\"u}semeister}, {Bucciarelli}, {Busonero}, {Buzzi}, {Cancelliere},
  {Carlucci}, {Charlot}, {Cheek}, {Crosta}, {Crowley}, {de Bruijne}, {de
  Felice}, {Drimmel}, {Esquej}, {Fienga}, {Fraile}, {Gai}, {Garralda},
  {Gonz{\'a}lez-Vidal}, {Guerra}, {Hauser}, {Hofmann}, {Holl}, {Jordan},
  {Lattanzi}, {Lenhardt}, {Liao}, {Licata}, {Lister}, {L{\"o}ffler},
  {Marchant}, {Martin-Fleitas}, {Messineo}, {Mignard}, {Morbidelli}, {Poggio},
  {Riva}, {Rowell}, {Salguero}, {Sarasso}, {Sciacca}, {Siddiqui}, {Smart},
  {Spagna}, {Steele}, {Taris}, {Torra}, {van Elteren}, {van Reeven}, \&
  {Vecchiato}}]{lindegren2018}
{Lindegren}, L., {Hern{\'a}ndez}, J., {Bombrun}, A., {et~al.} 2018, \aap, 616,
  A2, \dodoi{10.1051/0004-6361/201832727}

\bibitem[{{Lopez} \& {Fortney}(2013)}]{lopez2013}
{Lopez}, E.~D., \& {Fortney}, J.~J. 2013, \apj, 776, 2,
  \dodoi{10.1088/0004-637X/776/1/2}

\bibitem[{{Mann} {et~al.}(2012){Mann}, {Gaidos}, {L{\'e}pine}, \&
  {Hilton}}]{mann2012}
{Mann}, A.~W., {Gaidos}, E., {L{\'e}pine}, S., \& {Hilton}, E.~J. 2012, \apj,
  753, 90, \dodoi{10.1088/0004-637X/753/1/90}

\bibitem[{{Martinez} {et~al.}(2019){Martinez}, {Cunha}, {Ghezzi}, \&
  {Smith}}]{martinez2019}
{Martinez}, C.~F., {Cunha}, K., {Ghezzi}, L., \& {Smith}, V.~V. 2019, \apj,
  875, 29, \dodoi{10.3847/1538-4357/ab0d93}

\bibitem[{{Mathur} {et~al.}(2017){Mathur}, {Huber}, {Batalha}, {Ciardi},
  {Bastien}, {Bieryla}, {Buchhave}, {Cochran}, {Endl}, {Esquerdo}, {Furlan},
  {Howard}, {Howell}, {Isaacson}, {Latham}, {MacQueen}, \&
  {Silva}}]{mathur2017}
{Mathur}, S., {Huber}, D., {Batalha}, N.~M., {et~al.} 2017, \apjs, 229, 30,
  \dodoi{10.3847/1538-4365/229/2/30}

\bibitem[{{Mayor} {et~al.}(2011){Mayor}, {Marmier}, {Lovis}, {Udry},
  {S{\'e}gransan}, {Pepe}, {Benz}, {Bertaux}, {Bouchy}, {Dumusque}, {Lo Curto},
  {Mordasini}, {Queloz}, \& {Santos}}]{mayor2011}
{Mayor}, M., {Marmier}, M., {Lovis}, C., {et~al.} 2011, arXiv e-prints,
  arXiv:1109.2497.
\newblock \doarXiv{1109.2497}

\bibitem[{{Mazeh} {et~al.}(2016){Mazeh}, {Holczer}, \& {Faigler}}]{mazeh2016}
{Mazeh}, T., {Holczer}, T., \& {Faigler}, S. 2016, \aap, 589, A75,
  \dodoi{10.1051/0004-6361/201528065}

\bibitem[{{McDonald} {et~al.}(2019){McDonald}, {Kreidberg}, \&
  {Lopez}}]{mcdonald2019}
{McDonald}, G.~D., {Kreidberg}, L., \& {Lopez}, E. 2019, \apj, 876, 22,
  \dodoi{10.3847/1538-4357/ab109510.48550/arXiv.2105.00142}

\bibitem[{{Morton} {et~al.}(2016){Morton}, {Bryson}, {Coughlin}, {Rowe},
  {Ravichandran}, {Petigura}, {Haas}, \& {Batalha}}]{morton2016}
{Morton}, T.~D., {Bryson}, S.~T., {Coughlin}, J.~L., {et~al.} 2016, \apj, 822,
  86, \dodoi{10.3847/0004-637X/822/2/86}

\bibitem[{{Morton} \& {Swift}(2014)}]{morton2014}
{Morton}, T.~D., \& {Swift}, J. 2014, \apj, 791, 10,
  \dodoi{10.1088/0004-637X/791/1/10}

\bibitem[{{Mulders} {et~al.}(2015{\natexlab{a}}){Mulders}, {Pascucci}, \&
  {Apai}}]{mulders2015}
{Mulders}, G.~D., {Pascucci}, I., \& {Apai}, D. 2015{\natexlab{a}}, \apj, 798,
  112, \dodoi{10.1088/0004-637X/798/2/112}

\bibitem[{{Mulders} {et~al.}(2015{\natexlab{b}}){Mulders}, {Pascucci}, \&
  {Apai}}]{mulders2015c}
---. 2015{\natexlab{b}}, \apj, 814, 130, \dodoi{10.1088/0004-637X/814/2/130}

\bibitem[{{Mulders} {et~al.}(2015{\natexlab{c}}){Mulders}, {Pascucci}, \&
  {Apai}}]{mulders2015a}
---. 2015{\natexlab{c}}, \apj, 798, 112, \dodoi{10.1088/0004-637X/798/2/112}

\bibitem[{{Mulders} {et~al.}(2016){Mulders}, {Pascucci}, {Apai}, {Frasca}, \&
  {Molenda-{\.Z}akowicz}}]{mulders2016}
{Mulders}, G.~D., {Pascucci}, I., {Apai}, D., {Frasca}, A., \&
  {Molenda-{\.Z}akowicz}, J. 2016, \aj, 152, 187,
  \dodoi{10.3847/0004-6256/152/6/187}

\bibitem[{{Mullally} {et~al.}(2016){Mullally}, {Coughlin}, {Thompson},
  {Christiansen}, {Burke}, {Clarke}, \& {Haas}}]{mullally2016}
{Mullally}, F., {Coughlin}, J.~L., {Thompson}, S.~E., {et~al.} 2016, \pasp,
  128, 074502, \dodoi{10.1088/1538-3873/128/965/074502}

\bibitem[{{Narang} {et~al.}(2018){Narang}, {Manoj}, {Furlan}, {Mordasini},
  {Henning}, {Mathew}, {Banyal}, \& {Sivarani}}]{narang2018}
{Narang}, M., {Manoj}, P., {Furlan}, E., {et~al.} 2018, \aj, 156, 221,
  \dodoi{10.3847/1538-3881/aae391}

\bibitem[{{Owen} \& {Lai}(2018)}]{owen2018}
{Owen}, J.~E., \& {Lai}, D. 2018, \mnras, 479, 5012,
  \dodoi{10.1093/mnras/sty1760}

\bibitem[{{Owen} \& {Wu}(2017)}]{owen2017}
{Owen}, J.~E., \& {Wu}, Y. 2017, \apj, 847, 29,
  \dodoi{10.3847/1538-4357/aa890a}

\bibitem[{{Pecaut} \& {Mamajek}(2013)}]{pecaut2013}
{Pecaut}, M.~J., \& {Mamajek}, E.~E. 2013, \apjs, 208, 9,
  \dodoi{10.1088/0067-0049/208/1/9}

\bibitem[{{Petigura} {et~al.}(2013){Petigura}, {Howard}, \&
  {Marcy}}]{petigura2013}
{Petigura}, E.~A., {Howard}, A.~W., \& {Marcy}, G.~W. 2013, Proceedings of the
  National Academy of Science, 110, 19273, \dodoi{10.1073/pnas.1319909110}

\bibitem[{{Petigura} {et~al.}(2017){Petigura}, {Howard}, {Marcy}, {Johnson},
  {Isaacson}, {Cargile}, {Hebb}, {Fulton}, {Weiss}, {Morton}, {Winn}, {Rogers},
  {Sinukoff}, {Hirsch}, \& {Crossfield}}]{petigura2017}
{Petigura}, E.~A., {Howard}, A.~W., {Marcy}, G.~W., {et~al.} 2017, \aj, 154,
  107, \dodoi{10.3847/1538-3881/aa80de}

\bibitem[{{Petigura} {et~al.}(2018){Petigura}, {Marcy}, {Winn}, {Weiss},
  {Fulton}, {Howard}, {Sinukoff}, {Isaacson}, {Morton}, \&
  {Johnson}}]{petigura2018}
{Petigura}, E.~A., {Marcy}, G.~W., {Winn}, J.~N., {et~al.} 2018, \aj, 155, 89,
  \dodoi{10.3847/1538-3881/aaa54c}

\bibitem[{{Petigura} {et~al.}(2022){Petigura}, {Rogers}, {Isaacson}, {Owen},
  {Kraus}, {Winn}, {MacDougall}, {Howard}, {Fulton}, {Kosiarek}, {Weiss},
  {Behmard}, \& {Blunt}}]{petigura2022}
{Petigura}, E.~A., {Rogers}, J.~G., {Isaacson}, H., {et~al.} 2022, \aj, 163,
  179, \dodoi{10.3847/1538-3881/ac51e3}

\bibitem[{{Rogers} \& {Owen}(2021)}]{rogers2021}
{Rogers}, J.~G., \& {Owen}, J.~E. 2021, \mnras, 503, 1526,
  \dodoi{10.1093/mnras/stab529}

\bibitem[{Schwarz(1978)}]{schwarz1978}
Schwarz, G. 1978, The Annals of Statistics, 6, 461 ,
  \dodoi{10.1214/aos/1176344136}

\bibitem[{{Silburt} {et~al.}(2015){Silburt}, {Gaidos}, \& {Wu}}]{silburt2015}
{Silburt}, A., {Gaidos}, E., \& {Wu}, Y. 2015, \apj, 799, 180,
  \dodoi{10.1088/0004-637X/799/2/180}

\bibitem[{{Silverman}(1986)}]{silverman1986}
{Silverman}, B.~W. 1986, Density estimation for statistics and data analysis
  (Chapman and Hall)

\bibitem[{{Sullivan} {et~al.}(2022){Sullivan}, {Kraus}, \&
  {Mann}}]{sullivan2022}
{Sullivan}, K., {Kraus}, A.~L., \& {Mann}, A.~W. 2022, \apj, 935, 141,
  \dodoi{10.3847/1538-4357/ac7be9}

\bibitem[{{Swift} {et~al.}(2013){Swift}, {Johnson}, {Morton}, {Crepp},
  {Montet}, {Fabrycky}, \& {Muirhead}}]{swift2013}
{Swift}, J.~J., {Johnson}, J.~A., {Morton}, T.~D., {et~al.} 2013, \apj, 764,
  105, \dodoi{10.1088/0004-637X/764/1/105}

\bibitem[{{Thompson} {et~al.}(2015){Thompson}, {Mullally}, {Coughlin},
  {Christiansen}, {Henze}, {Haas}, \& {Burke}}]{thompson2015}
{Thompson}, S.~E., {Mullally}, F., {Coughlin}, J., {et~al.} 2015, \apj, 812,
  46, \dodoi{10.1088/0004-637X/812/1/4610.48550/arXiv.1509.00041}

\bibitem[{{Thompson} {et~al.}(2018){Thompson}, {Coughlin}, {Hoffman},
  {Mullally}, {Christiansen}, {Burke}, {Bryson}, {Batalha}, {Haas},
  {Catanzarite}, {Rowe}, {Barentsen}, {Caldwell}, {Clarke}, {Jenkins}, {Li},
  {Latham}, {Lissauer}, {Mathur}, {Morris}, {Seader}, {Smith}, {Klaus},
  {Twicken}, {Van Cleve}, {Wohler}, {Akeson}, {Ciardi}, {Cochran}, {Henze},
  {Howell}, {Huber}, {Pr{\v{s}}a}, {Ram{\'\i}rez}, {Morton}, {Barclay},
  {Campbell}, {Chaplin}, {Charbonneau}, {Christensen-Dalsgaard}, {Dotson},
  {Doyle}, {Dunham}, {Dupree}, {Ford}, {Geary}, {Girouard}, {Isaacson},
  {Kjeldsen}, {Quintana}, {Ragozzine}, {Shabram}, {Shporer}, {Silva Aguirre},
  {Steffen}, {Still}, {Tenenbaum}, {Welsh}, {Wolfgang}, {Zamudio}, {Koch}, \&
  {Borucki}}]{thompson2018}
{Thompson}, S.~E., {Coughlin}, J.~L., {Hoffman}, K., {et~al.} 2018, \apjs, 235,
  38, \dodoi{10.3847/1538-4365/aab4f9}

\bibitem[{Towns {et~al.}(2014)Towns, Cockerill, Dahan, Foster, Gaither,
  Grimshaw, Hazlewood, Lathrop, Lifka, Peterson, Roskies, Scott, \&
  Wilkins-Diehr}]{xsede}
Towns, J., Cockerill, T., Dahan, M., {et~al.} 2014, Computing in Science \&
  Engineering, 16, 62, \dodoi{10.1109/MCSE.2014.80}

\bibitem[{{Traub}(2012)}]{traub2012}
{Traub}, W.~A. 2012, \apj, 745, 20, \dodoi{10.1088/0004-637X/745/1/20}

\bibitem[{van~der Walt {et~al.}(2011)van~der Walt, Colbert, \&
  Varoquaux}]{numpy}
van~der Walt, S., Colbert, S.~C., \& Varoquaux, G. 2011, Computing in Science
  Engineering, 13, 22, \dodoi{10.1109/MCSE.2011.37}

\bibitem[{{Van Eylen} {et~al.}(2018){Van Eylen}, {Agentoft}, {Lundkvist},
  {Kjeldsen}, {Owen}, {Fulton}, {Petigura}, \& {Snellen}}]{vaneylan2018}
{Van Eylen}, V., {Agentoft}, C., {Lundkvist}, M.~S., {et~al.} 2018, \mnras,
  479, 4786, \dodoi{10.1093/mnras/sty1783}

\bibitem[{Virtanen {et~al.}(2020)Virtanen, Gommers, Oliphant, Haberland, Reddy,
  Cournapeau, Burovski, Peterson, Weckesser, Bright, {van der Walt}, Brett,
  Wilson, Millman, Mayorov, Nelson, Jones, Kern, Larson, Carey, Polat, Feng,
  Moore, {VanderPlas}, Laxalde, Perktold, Cimrman, Henriksen, Quintero, Harris,
  Archibald, Ribeiro, Pedregosa, {van Mulbregt}, \& {SciPy 1.0
  Contributors}}]{scipy}
Virtanen, P., Gommers, R., Oliphant, T.~E., {et~al.} 2020, Nature Methods, 17,
  261, \dodoi{10.1038/s41592-019-0686-2}

\bibitem[{{Wang} {et~al.}(2015){Wang}, {Fischer}, {Horch}, \&
  {Huang}}]{wang2015}
{Wang}, J., {Fischer}, D.~A., {Horch}, E.~P., \& {Huang}, X. 2015, \apj, 799,
  229, \dodoi{10.1088/0004-637X/799/2/229}

\bibitem[{{Weiss} \& {Marcy}(2014)}]{weiss2014}
{Weiss}, L.~M., \& {Marcy}, G.~W. 2014, \apjl, 783, L6,
  \dodoi{10.1088/2041-8205/783/1/L6}

\bibitem[{{W}es {M}c{K}inney(2010)}]{pandas}
{W}es {M}c{K}inney. 2010, in {P}roceedings of the 9th {P}ython in {S}cience
  {C}onference, ed. {S}t\'efan van~der {W}alt \& {J}arrod {M}illman, 56 -- 61,
  \dodoi{10.25080/Majora-92bf1922-00a}

\bibitem[{{Wittenmyer} {et~al.}(2020){Wittenmyer}, {Wang}, {Horner}, {Butler},
  {Tinney}, {Carter}, {Wright}, {Jones}, {Bailey}, {O'Toole}, \&
  {Johns}}]{wittenmeyer2020}
{Wittenmyer}, R.~A., {Wang}, S., {Horner}, J., {et~al.} 2020, \mnras, 492, 377,
  \dodoi{10.1093/mnras/stz3436}

\bibitem[{{Wright} {et~al.}(2012){Wright}, {Marcy}, {Howard}, {Johnson},
  {Morton}, \& {Fischer}}]{wright2012}
{Wright}, J.~T., {Marcy}, G.~W., {Howard}, A.~W., {et~al.} 2012, \apj, 753,
  160, \dodoi{10.1088/0004-637X/753/2/160}

\bibitem[{{Wu}(2019)}]{wu2019}
{Wu}, Y. 2019, \apj, 874, 91, \dodoi{10.3847/1538-4357/ab06f8}

\bibitem[{{Yang} {et~al.}(2020){Yang}, {Xie}, \& {Zhou}}]{yang2020}
{Yang}, J.-Y., {Xie}, J.-W., \& {Zhou}, J.-L. 2020, \aj, 159, 164,
  \dodoi{10.3847/1538-3881/ab7373}

\bibitem[{{Youdin}(2011)}]{youdin2011}
{Youdin}, A.~N. 2011, \apj, 742, 38, \dodoi{10.1088/0004-637X/742/1/38}

\bibitem[{{Zhu} {et~al.}(2018){Zhu}, {Petrovich}, {Wu}, {Dong}, \&
  {Xie}}]{zhu2018}
{Zhu}, W., {Petrovich}, C., {Wu}, Y., {Dong}, S., \& {Xie}, J. 2018, \apj, 860,
  101, \dodoi{10.3847/1538-4357/aac6d5}

\bibitem[{{Zink} {et~al.}(2019){Zink}, {Christiansen}, \& {Hansen}}]{zink2019}
{Zink}, J.~K., {Christiansen}, J.~L., \& {Hansen}, B. M.~S. 2019, \mnras, 483,
  4479, \dodoi{10.1093/mnras/sty3463}

\end{thebibliography}
\bibliographystyle{aasjournal}

\end{document}